\documentclass{article}
\usepackage{pgfplots}
\usepackage{lipsum}
\usepackage{caption}
\usepackage{subcaption}
\usepackage{authblk}
\usepackage[english]{babel}
\usepackage{amsfonts}
\usepackage{amsmath}
\usepackage[top=2cm, bottom=2cm, left=2cm, right=2cm]{geometry}
\usepackage{authblk}
\usepackage{fancyhdr}
\usepackage{multicol}
\usepackage{bbold}
\usepackage{epsfig}
\usepackage{graphics}
\usepackage{epsfig}
\usepackage{lipsum}
\usepackage{fancyhdr}
\usepackage{booktabs}

\usepackage{wrapfig}
\usepackage[export]{adjustbox}
\usepackage{rotating}
\usepackage{amsmath}
\usepackage{cite}
\usepackage[colorlinks=true,linkcolor=blue]{hyperref} 
\usepackage{cleveref}

\renewenvironment{abstract}{
	
	\hfill\begin{minipage}{0.95\textwidth}
		\rule{\textwidth}{1pt}}
	{\par\noindent\rule{\textwidth}{1pt}\end{minipage}}

\makeatletter
\usepackage{orcidlink}
\makeatother
\begin{document}


\title{Geometry and quantum brachistochrone analysis of multiple entangled spin-1/2 particles under all-range Ising interaction}

	\author[1,2]{\textbf{B. Amghar *\normalsize\orcidlink{0000-0002-5284-3240}}}
	\author[2,3]{\textbf{M. Yachi}}
 \author[4]{\textbf{M. Amghar}}
 \author[5]{\textbf{M. Almousa}}
\author[6,7]{\textbf{A. A. Abd El-Latif}}
 \author[2,6,8]{\textbf{ A. Slaoui *\normalsize\orcidlink{0000-0002-5284-3240}}}
	\affil[1]{\small Laboratory LPNAMME, Laser Physics Group, Department of Physics, Faculty of Sciences, Chouaïb Doukkali University, El Jadida, Morocco.}
	\affil[2]{\small LPHE-Modeling and Simulation, Faculty of Sciences, Mohammed V University in Rabat, Rabat, Morocco.}
 \affil[3]{\small Laboratory of Innovation in Science, Technology and Modeling, Faculty of Sciences, Chouaïb Doukkali University, El Jadida, Morocco}
 \affil[4]{\small LPTHE, Department of Physics, Faculty of Sciences, Ibn Zohr University, Agadir, Morocco.}
\affil[5]{\small Department of Information Technology, College of Computer and Information Sciences, Princess Nourah bint Abdulrahman University, Riyadh 11671, Saudi Arabia.}
\affil[6]{\small EIAS Data Science Lab, College of Computer and Information Sciences, and Center of Excellence in Quantum and Intelligent Computing, Prince Sultan University, Riyadh 11586, Saudi Arabia.}
\affil[7]{\small Department of Mathematics and Computer Science, Faculty of Science, Menoufia University, Shebin El-Koom 32511, Egypt.}
\affil[8]{\small Centre of Physics and Mathematics, CPM, Faculty of Sciences, Mohammed V University in Rabat, Rabat, Morocco.\\ *Contact author: b.brahim.amghar@gmail.com, abdallah.slaoui@um5s.net.ma}

	\maketitle
	
	\begin{center}
		\textbf{Abstract}
	\end{center}
	\begin{abstract}

We present a unified geometric and dynamical framework for a physical system consisting of $n$ spin-$1/2$ particles with all-range Ising interaction. Using the Fubini-Study formalism, we derive the metric tensor of the associated quantum state manifold and compute the corresponding Riemann curvature. Our analysis reveals that the system evolves over a smooth, compact, two-dimensional manifold with spherical topology and a dumbbell-like structure shaped by collective spin interactions. We further investigate the influence of the geometry and topology of the resulting state space on the behavior of geometric and topological phases acquired by the system. We explore how this curvature constrains the system's dynamical behavior, including its evolution speed and Fubini-Study distance between the quantum states. Within this geometric framework, we address the quantum brachistochrone problem and derive the minimal time required for optimal evolution, a result useful for time-efficient quantum circuit design. Subsequently, we explore the role of entanglement in shaping the state space geometry, modulating geometric phase, and controlling evolution speed and brachistochrone time. Our results reveal that entanglement enhances dynamics up to a critical threshold, beyond which geometric constraints begin to hinder evolution. Moreover, entanglement induces critical shifts in the geometric phase, making it a sensitive indicator of entanglement levels and a practical tool for steering quantum evolution. This approach offers valuable guidance for developing quantum technologies that require time-efficient control strategies rooted in the geometry of quantum state space.\par
\end{abstract}
	
	\vspace{0.5cm}
	
\section{Introduction}
In recent years, the integration of geometric approaches into quantum physics has received increasing attention, as geometry offers a powerful lens for interpreting the dynamical properties of quantum systems. By incorporating geometric notions into the foundations of quantum theory, researchers have developed robust frameworks for analyzing solvable quantum models and uncovering deeper physical insights. This approach not only enriches our conceptual understanding of quantum mechanics but also contributes meaningfully to the broader field of information geometry \cite{Kibble1979,Anandan1991,Ashtekar1999,Brody2001}. Central to this perspective is the idea that the space of quantum states naturally forms a Kähler manifold, an elegant structure that supports a rigorous treatment of quantum evolution and the intricate relationships between states \cite{Provost1980,Zhang1995,Botero2003,Rawnsley1977}. Recent studies have emphasized that the geometrical characteristics of the space of quantum states help quantify, to a large extent, dynamical properties. For example, the Fubini-Study distance, a natural metric on projective Hilbert space $\mathbb{C}P^n$, has been related to the energy uncertainty of a system, thereby linking the geometry of evolutionary paths to the rate at which quantum states change \cite{Anandan1990}. In the broader context of quantum dynamics, the quantum speed limit, defining the minimal time required for a system to evolve between two distinguishable states, is expressed through geometric measures such as the Bures distance between the mixed states \cite{Deffner2013}. Moreover, geometric methods have often proven effective in addressing the quantum brachistochrone problem, which involves identifying the Hamiltonian that yields the fastest possible evolution between two states \cite{Ammghar2020, Bender2007, Frydryszak2008}. In quantum computing, the geometry of quantum state space has been effectively utilized to study the design of efficient quantum circuits. For systems of $n$ qutrits, it has been shown that the optimal quantum circuits correspond closely to the shortest path between two points within the specific curved geometry of $SU(3^n)$, analogous to the role of $SU(2^n)$ in qubit-based systems \cite{Li2013,Nielsen2006}. For further studies that highlight the relevance of geometry in understanding quantum system dynamics, see refs. \cite{Dowling2008, Deffner2010, Pires2015, Pires2016}.\par
Currently, geometric quantum mechanics serves as the foundational framework for the geometric formulation of quantum theory. It has established itself as a cornerstone of information geometry. Within this paradigm, quantum phenomena are examined through a geometric perspective \cite{Horodecki2009,Elfakir2024,Amico2008,Slaoui2023,Amghar2020}. A notable example of this approach is quantum entanglement, which is recognized as one of the most powerful resources in modern quantum science. It drives progress in quantum computing, secure communication, high-precision sensing, and advanced measurement techniques. While it is often studied using algebraic or statistical tools, there is growing interest in viewing entanglement through a geometric lens, an approach that offers conceptual clarity as well as practical advantages. Geometric frameworks help to visualize and better understand the structure of entangled states, especially in systems involving many particles \cite{k1}. The metrics, such as the Euclidean or trace distances, quantify the deviation of an entangled state from separability, offering a tangible means of evaluating its quantum correlations \cite{k2}. These tools also play an important role in constructing entanglement witnesses, optimizing purification schemes, and analyzing the shape and boundaries of the quantum state space \cite{Bertlmann2002}. From a mathematical standpoint, techniques such as Hopf fibrations and projective geometry offer powerful ways to represent entanglement, linking it to symmetry and topological properties \cite{Mosseri2006}. Crucially, these discoveries are not limited to theory; they are relevant in real experiments with trapped ions, superconducting circuits, and cold atoms \cite{c8,c9,c10,bha}. For additional ideas that illuminate the geometric aspects of entanglement, please consult refs. \cite{Mosseri2001,Amghar2023,Verstraete2002}.\par
An influential concept that has garnered considerable attention in quantum physics is the geometric phase, a fascinating characteristic that arises during the evolution of quantum systems. This phase is essential for understanding the influence of geometric properties on quantum dynamics, revealing how the path taken by a system in parameter space can lead to distinct physical outcomes. The study of the geometric phase has important implications across diverse areas, such as quantum computation and quantum control, thereby deepening our comprehension of fundamental quantum phenomena.
\cite{Berry1984,Aharonov1987,Slaoui2024,Anandan1992}. It is often interpreted as the holonomy accumulated by the quantum state vector during parallel transport along its evolution path  \cite{Andersson2019,Demler1999}. Presently, the geometric phase has been strongly linked to intrinsic properties of quantum state manifolds. In particular, it has been shown to correspond to the line integral of the Berry–Simon connection over the Fubini–Study path length, which measures the quantum distance between states in the relevant projective Hilbert space \cite{Botero2003, Samuel1988}. On a practical level, numerous works have highlighted the importance of the geometric phase in advancing quantum information technologies, especially its role in implementing reliable operations in quantum circuits, cryptographic systems, and other frontier applications \cite{Wang2001, Kleipler2018, Kumar2005}.  For instance, conditional phase gates have been successfully demonstrated in nuclear magnetic resonance experiments \cite{Jones2000} and with trapped ion systems \cite{Duan2001}.  A broader overview of its applications in quantum information can be found in works such as \cite{Oxman2011,Khoury2014,Johansson2012,Vedral2003}.\par
The present study is motivated by the insightful work of Krynytskyi and Kuzmak, as reported in Ref. \cite{Krynytskyi2019}, who laid the foundations for a geometric understanding of the evolution of a spin-$s$ system with the long-range $zz$-type
Ising interaction. In particular, the study derives the Fubini-Study metric, identifies a spherical topology of the quantum state space, and relates the curvature of the state space to the evolution rate. However, this approach remains focused on global aspects of the considered system. \par Our main purpose extends and enriches this approach by developing a unified framework, both geometric and dynamical, for a multipartite system composed of $n$ spin-$1/2$ particles with all-range Ising interaction. In addition to precisely characterizing the metric, curvature, and topology of the state space, we provide intuitive
discussions of how the topology and the geometry of the resulting quantum state space affect the quantum entanglement, evolution speed, Fubini-Study distance and optimal evolution time (quantum brachistochrone). Then, we investigate the critical role of entanglement in limiting or accelerating the dynamics, highlighting the threshold beyond which geometric effects restrict the evolution. Moreover, we show that entanglement induces critical shifts in the geometric phase, making it a sensitive indicator of entanglement levels and a practical tool for steering quantum evolution. This interplay between entanglement and geometry reveals fundamental limits of quantum control and offers valuable guidance for designing time-efficient quantum protocols.\par
The current paper is organized as follows. In section \ref{sec2}, by carrying out the dynamics of $n$ spin-$1/2$ particles with all-range Ising-type interaction, we derive the metric tensor of the associated quantum state manifold and compute the corresponding Riemann curvature, revealing the topology and inherent structure of this manifold. We investigate the influence of the manifold’s curvature and topology on the emergence of geometric, dynamic, and topological phases during system evolution, in section \ref{sec4}. In section \ref{sec3}, we address the quantum brachistochrone problem within this geometric framework, analyzing the evolution speed and the corresponding Fubini–Study distance to determine the minimal time required for optimal state evolution. In section \ref{sec5}, by reducing the entire system to two spin-1/2 particles, we provide a detailed description of exchanged quantum entanglement from two different perspectives:  the first pertains to the geometric side and explores the entanglement influence on the various geometric features, including the state space geometry, R-curvature, and geometric phase. The second pertains to the dynamic side and investigates the influence of the entanglement on the evolution rate and the relevant FS-distance (Fubini-Study distance). Additionally, we address the brachistochrone issue in terms of the entanglement degree exchanged between the two spins. In Section \ref{sec6}, we conclude the paper with a summary.
\section{Quantum evolution and the quantum state space of $n$ spin-$1/2$ particles}\label{sec2}
\subsection{Ising model and unitary dynamics} 
\begin{figure}[htbphtbp]
		\centering \includegraphics[width=10cm,height=4cm]{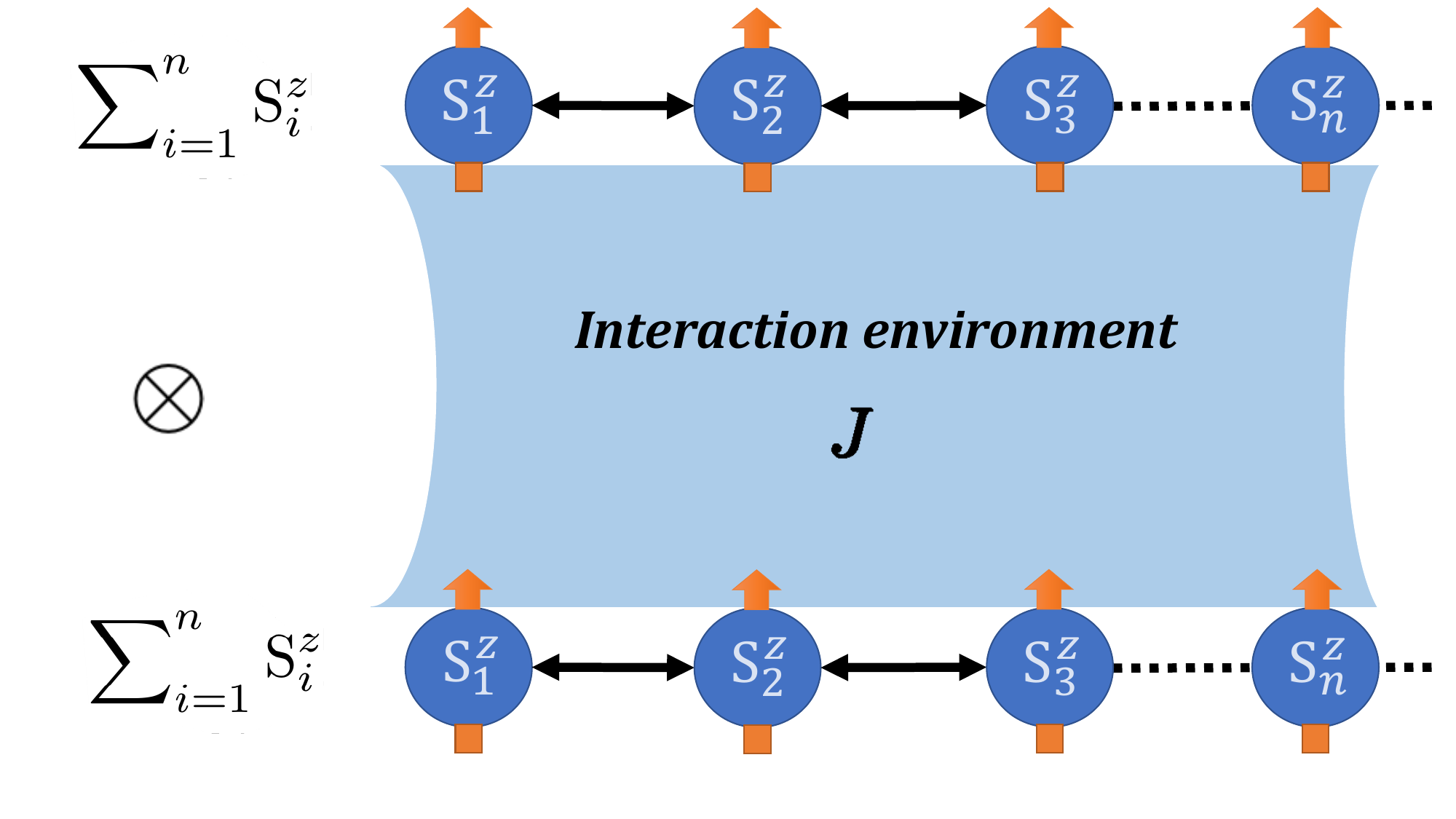}
		\caption{{Illustrative scheme of two coupled $n$-spin-$1/2$ chains under all-range Ising interaction.}}
		\label{Nspin}
	\end{figure}

In this work, we study a multipartite system consisting of $n$ spin-1/2 particles with all-range Ising model interaction {(see Fig \eqref{Nspin})}. 
    
The Hamiltonian governing this system reads
\begin{equation}\label{1}
 	{H}={J}\left[\sum\nolimits\nolimits_{i = 1}^n \mathrm{S}^z_i\right] ^2,
 \end{equation} 
where ${J}$ denotes the coupling parameter associated with the Ising interaction, while $\mathrm{S}_k^z$ represents the component along the z-axis of the spin operator $\boldsymbol{{\mathrm{S}}}_i=(\mathrm{S}_i^x,\mathrm{S}_i^y,\mathrm{S}_i^z)^T$ corresponding to $i$-$th$ spin$-1/2$. It verifies the eigenvalue equation
\begin{equation}\label{2}
\mathrm{S}_i^z\left| {{\mathrm{m}_i}} \right\rangle  ={\mathrm{m}_i} \left| {{\mathrm{m}_i}} \right\rangle,
\end{equation}
with $S_i^{\kappa}=\frac{\hbar}{2}\sigma_i^{\kappa}$ and $\sigma_i^{\kappa}$\, $(\kappa=x,y,z)$ are the Pauli matrices, $\mathrm{m}_{i}=\pm \hbar/2$ are the possible values obtained by projecting the $i$-$th$ spin along the $z$-axis, and $\left| {{\mathrm{m}_i}} \right\rangle $ denote the corresponding eigenstates. It is noteworthy that the components of spin$-1/2$ operators $\mathrm{S}_i^x,\,\mathrm{S}_i^y,$ and $\mathrm{S}_i^z$ satisfy the algebraic structure related to the Lie group $SU(2)$
\begin{equation}
\left[ {\mathrm{S}_i^a ,\mathrm{S}_j^b } \right] = i{\delta _{ij}}\sum\nolimits\nolimits_{c  = x,y,z}\epsilon^{abc} {\mathrm{S}_i^c },
\end{equation}
with $\delta _{ij}$ and $\epsilon^{ijk}$ are, respectively, the Kronecker and Levi-Civita symbols. Remark that in the case of an even number of spins, the above Hamiltonian possesses $(n/2 + 1)$ eigenvalues, whilst for an odd number, it has $(n + 1)/2$ eigenvalues. The eigenvalues and their associated eigenstates are explicitly presented as follows
	\begin{equation}\label{3}
\begin{array}{cc}
\frac{J\hbar^2}{4} n^2 & |\frac{1}{2}, \frac{1}{2}, \frac{1}{2}, \ldots, \frac{1}{2}, \frac{1}{2}\rangle,|-\frac{1}{2}, -\frac{1}{2}, -\frac{1}{2}, \ldots, -\frac{1}{2}, -\frac{1}{2}\rangle ; \\[12px]
\frac{{J}\hbar^2}{4}(n-2)^2 & |-\frac{1}{2}, \frac{1}{2}, \frac{1}{2}, \ldots\frac{1}{2}, \frac{1}{2}\rangle,|\frac{1}{2}, -\frac{1}{2}, \frac{1}{2}, \ldots, \frac{1}{2}, \frac{1}{2}\rangle, \ldots,|\frac{1}{2}, \frac{1}{2}, \frac{1}{2}, \ldots, \frac{1}{2}, -\frac{1}{2}\rangle, \\[6px]
& |\frac{1}{2}, -\frac{1}{2}, -\frac{1}{2}, \ldots, -\frac{1}{2}, -\frac{1}{2}\rangle,|-\frac{1}{2}, \frac{1}{2}, -\frac{1}{2}, \ldots, -\frac{1}{2}, -\frac{1}{2}\rangle, \ldots,|-\frac{1}{2}, -\frac{1}{2}, -\frac{1}{2}, \ldots, -\frac{1}{2}, \frac{1}{2}\rangle ; \\[12px]
\frac{{J}\hbar^2}{4}(n-4)^2 & |-\frac{1}{2}, -\frac{1}{2}, \frac{1}{2}, \ldots \frac{1}{2}, \frac{1}{2}\rangle,|-\frac{1}{2}, \frac{1}{2}, -\frac{1}{2}, \ldots \frac{1}{2}, \frac{1}{2}\rangle, \ldots,|\frac{1}{2}, \frac{1}{2}, \frac{1}{2}, \ldots, -\frac{1}{2}, -\frac{1}{2}\rangle, \\[6px]
& |\frac{1}{2}, \frac{1}{2}, -\frac{1}{2}, \ldots, -\frac{1}{2}, -\frac{1}{2}\rangle,|\frac{1}{2}, -\frac{1}{2}, \frac{1}{2}, \ldots, -\frac{1}{2}, -\frac{1}{2}\rangle, \ldots,|-\frac{1}{2}, -\frac{1}{2}, -\frac{1}{2}, \ldots, \frac{1}{2}, \frac{1}{2}\rangle ; \\[12px]
\ldots & \ldots
\end{array}
	\end{equation}
Considering all potential configurations of the spin states ($|\frac{1}{2}\rangle$ and $|-\frac{1}{2}\rangle$ ), we uncover that each eigenvalue $
{{J\hbar^2{{(n - 2p)}^2}} \mathord{\left/
 {\vphantom {{J{{(n - 2p)}^2}} 4}} \right.
 \kern-\nulldelimiterspace} 4}$ corresponds to $2\mathrm{C}_n^p$ eigenstates, where $\mathrm{C}$ denotes the binomial coefficient, whilst the index $p=0,...,n/2$ for even $n$ (number of particles)  and $p=0,...,(n-1)/{2}$ for odd $n$. We consider the evolution of the $n$ spin$-1/2$ system begins with the starting state
 \begin{equation}\label{a}
 |\psi_i\rangle=|\mathrm{S}\rangle ^{\otimes n},
 \end{equation}
	so that
		$$
			|\mathrm{S}\rangle=\cos {\theta  \mathord{\left/
 {\vphantom {\theta  2}} \right.
 \kern-\nulldelimiterspace} 2}\;|{{\frac{1}{2}}}\rangle+ \exp({i\varphi}) \sin {\theta  \mathord{\left/
 {\vphantom {\theta  2}} \right.
 \kern-\nulldelimiterspace} 2}|{{-\frac{1}{2}}}\rangle 
$$
represents an eigenstate of the spin-$1/2$ projection operator along the direction defined by the vector $$\textbf{v}=(\cos\varphi,\sin\varphi,\tan^{-1}\theta)\\\sin\theta,$$ where $\theta$ and $\varphi$ denote, respectively, the polar and azimuthal angles. In this respect, the initial state \eqref{a} can be rewritten using the binomial theorem as follows 
\begin{small}
\begin{equation}\label{b}
\left|\psi_i\right\rangle=\sum\nolimits_{p=0}^n \cos^{n}\frac{\theta}{2}\tan ^p\frac{\theta}{2}\exp({ip\varphi}) \sum\nolimits_{i_1<i_2<\ldots<i_p=1}^n \sigma_{i_1}^x \sigma_{i_2}^x \ldots \sigma_{i_p}^x|\frac{1}{2} \rangle ^{\otimes n},
		\end{equation}
\end{small}
where we set $\hbar= 1$. Now, to explore the geometrical, topological, and dynamical characteristics of the $n$ interacting spin-1/2 particles, we have to move this system, initially maintained in the starting state \eqref{b}, by applying the time evolution operator $\mathtt{P}(t)=\exp({-i{H}t})$. Thus, the evolving state of the considered system reads
		\begin{equation}\label{c}
			\left|\psi(t)\right\rangle=\sum\nolimits_{p=0}^n \cos^{n}\frac{\theta}{2}\tan ^p\frac{\theta}{2} \exp\left[-i\left(\frac{\chi(t)}{4}(n-2p)^2-p\varphi\right)\right]\sum\nolimits_{i_1<i_2<\ldots<i_p=1}^n \sigma_{i_1}^x \sigma_{i_2}^x \ldots \sigma_{i_p}^x|\frac{1}{2}, \frac{1}{2}, \ldots, \frac{1}{2}\rangle,
		\end{equation}
where we put $\chi(t)=Jt.$	This leads us to establish the evolving $n$ spin-1/2 state, which depends on the spherical angles $(\theta,\varphi)$ and the evolution duration $t$. Remark that the resulting state \eqref{c} satisfies the periodic Requirement $\left|\psi(\chi)\right\rangle =\left|\psi(\chi+2\pi)\right\rangle$ with respect to the dynamical variable $\chi$. This implies that $\chi\in[0,2\pi]$. Accordingly, we can conclude that the dynamics of the considered system occurs within a Bounded three-dimensional space. Now, it becomes natural to identify the physical space, i.e., the relevant quantum phase space, in which the system evolves by employing the Fubini–Study formalism, which provides the appropriate geometric framework for describing the quantum states of the system.

 \subsection{Intrinsic geometry of the resulting quantum state space}
\begin{figure}[h!]
		\centering \includegraphics[scale=0.135]{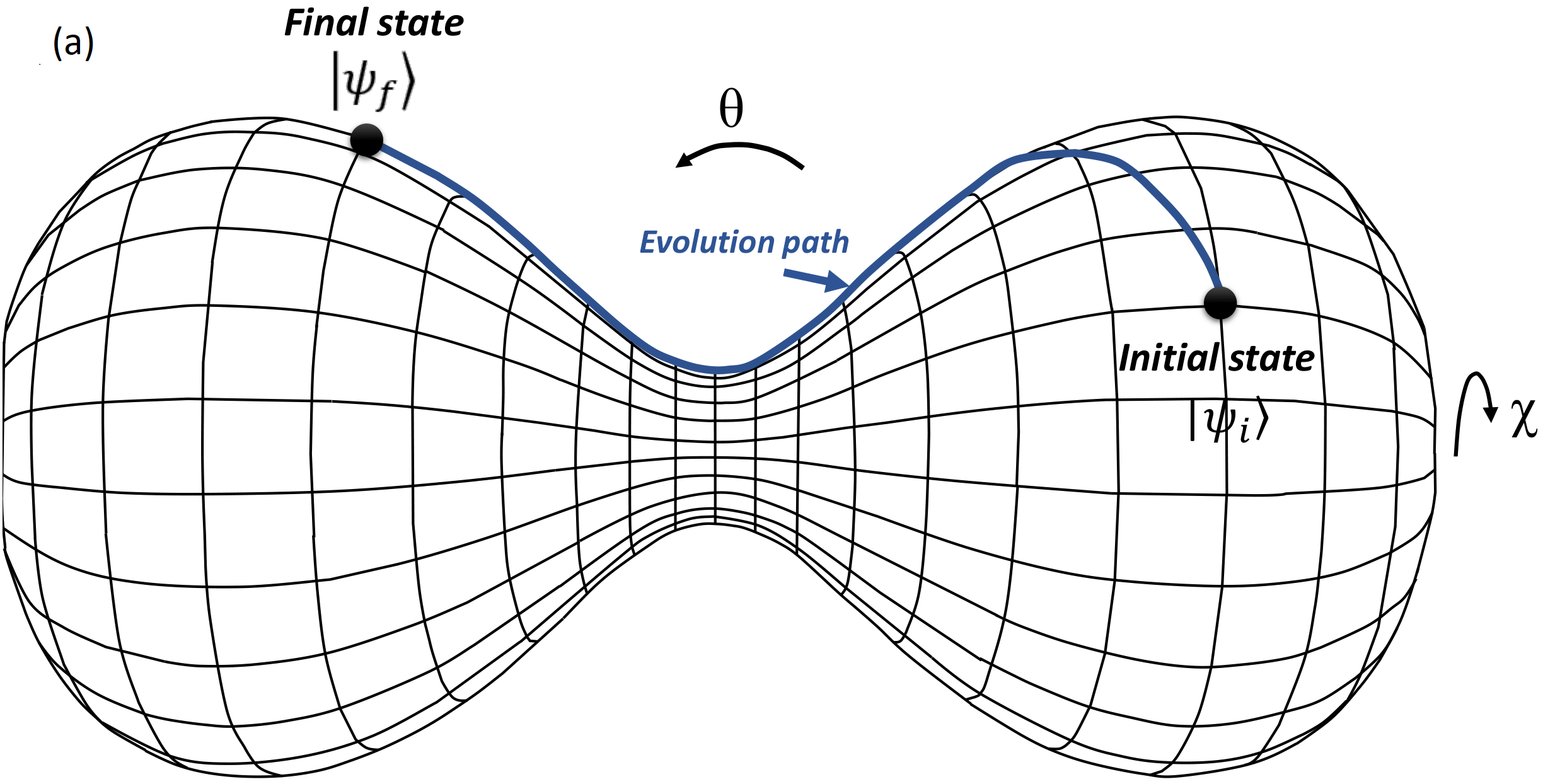}\includegraphics[scale=0.32]{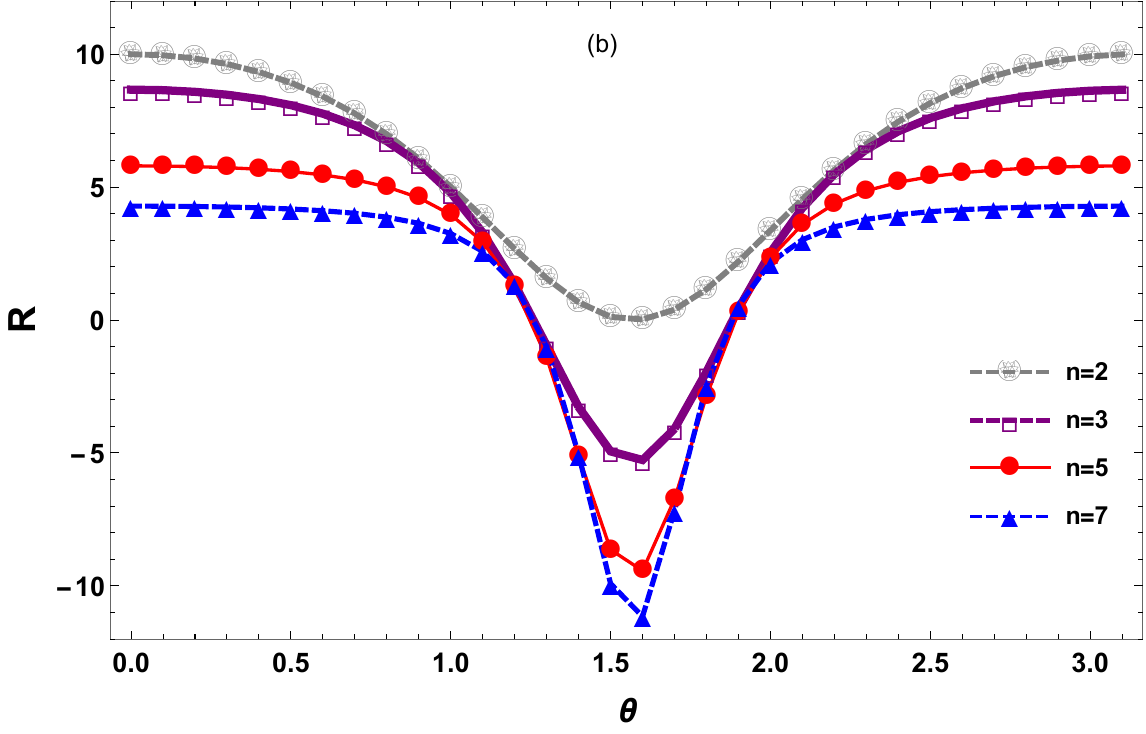}
		\caption{(a) Illustrative scheme of the quantum state space \eqref{f} of the $n$ spin-1/2 system under the all-range Ising model. (b) the behavior of the state space curvature \eqref{h} against the initial parameter $\theta$ for specific spin$-1/2$ numbers.}
		\label{Curve}
	\end{figure}

Having established the time-evolved state \eqref{c} of the $n$ spin-$1/2$ system, we now proceed to investigate the geometric structure of the corresponding quantum state space. This space comprises all accessible pure states generated by the system's unitary evolution. To characterize its geometry, we compute the FS-metric tensor (Fubini-Study metric tensor), which quantifies the infinitesimal distance $dS$ between two neighboring quantum states, $\left|\psi\left(\zeta^\alpha\right)\right\rangle$ and $\left|\psi\left(\zeta^\alpha+d \zeta^\alpha\right)\right\rangle$. The FS-distance is given by the expression \cite{Frydryszak2008}:
 		\begin{equation}\label{d}
 	dS^2=\mathrm{g}_{\alpha\beta} d \zeta^\alpha d \zeta^\beta,	
 		\end{equation}
where $\zeta^\alpha$ designate the physical parameters $\theta, \varphi$ and $\chi$ that characterize the $n$ spin-$1/2$ state \eqref{c} and $\mathrm{g}_{\alpha \beta}$ denote the metric tensor components, which are given by
\begin{equation}\label{e}
\mathrm{g}_{\alpha \beta}=\mathrm{Re}\left[\left\langle\psi_\alpha|\psi_\beta\right\rangle-\left\langle\psi_\alpha |\psi\right\rangle\left\langle\psi| \psi_\beta\right\rangle\right],
\end{equation}
where $\left|\psi_{\alpha,\beta}\right\rangle=\frac{\partial}{\partial \zeta^{\alpha,\beta}}|\psi\rangle$. Based on the definition \eqref{d}, we derive the explicit form of the FS-metric as
\begin{small}
\begin{align}\label{e}
 	dS^2= dS_i^2+ \frac{1}{8} n(n-1) \sin ^2 \theta\left[1+\left(2n-3\right) \cos ^2 \theta\right]d\chi^2 + \frac{1}{8} n(n-1)\sin 2\theta \sin \theta d\varphi d\chi,
 		\end{align}
\end{small}	
where
\begin{small}
\begin{equation}\label{a17}
dS_i^2= \frac{n}{4}\left(d \theta^2+\sin ^2 \theta d \varphi^2\right)
\end{equation}
\end{small}
This metric tensor captures the geometry of the quantum state space associated with the system's initial configuration. To better visualize this, consider the special case where $\chi=0$, meaning the system has not yet evolved. In this situation, the line element \eqref{e} shows that the state space simplifies to a sphere of radius $2\sqrt{n}$. Once the system begins to evolve under unitary dynamics, the quantum state explores a richer portion of projective Hilbert space, and the associated manifold of quantum states expands into a bounded, three-dimensional, curved space. Moreover, the components of the metric tensor \eqref{e} are independent of the azimuthal angle, $\varphi$. This rotational symmetry implies that all states with the same value of $\varphi$ share the same local geometry. Consequently, the effective geometry of the quantum state space for the $n$ spin-${1}/{2}$ particles reduces to a two-parameter manifold described by the polar angle $\theta$ and the evolution parameter $\chi$ {(see Fig. \ref{Curve}(a))}. The corresponding line element simplifies to

\begin{equation}\label{f}
 	dS^2=\frac{1}{4}\left[ n d \theta^2 +  n(n-1) \sin ^2 \theta\left(\frac{1}{2}+\left(n-\frac{3}{2}\right) \cos ^2 \theta\right)d\chi^2\right].
 		\end{equation}
which defines a smooth, curved, and bounded space containing all physical states of the system. It is worth noting that the obtained metric \eqref{f} is identical to that derived for a system of $n$ spin-$s$ particles governed by a long-range zz-type Ising interaction, as reported in \cite{Krynytskyi2019}. This coincidence arises because the two Hamiltonians differ only by local spin terms, which only contribute global phases and do not alter the geometry of the quantum evolution in the resulting quantum state space. As such, both models trace out identical state trajectories up to an overall phase and thus induce the same Fubini–Study metric structure. \par After identifying the Fubini-Study metric of the $n$ spin-$\frac{1}{2}$ system, it becomes essential to explore the associated curvature. While the metric describes local distances between quantum states, the Riemannian curvature offers a global measure of how the manifold bends. Moreover, it reveals how physical effects, like spin coupling and entanglement, shape the geometry of the state space. For this purpose, the relevant R-curvature (Riemann curvature) in relation to the FS-metric \eqref{f} under the form \cite{Kolodrubetz2013}
\begin{align}\label{g}
R=\frac{2}{{\sqrt {{{\rm{g}}_{\theta \theta }}{{\rm{g}}_{\chi \chi }}} }}\left[ {\frac{\partial }{{\partial \chi }}\left( {\Gamma _{\theta \theta }^\chi \sqrt {\frac{{{{\rm{g}}_{\chi \chi }}}}{{{{\rm{g}}_{\theta \theta }}}}} } \right)} \right.\left. { - \frac{\partial }{{\partial \theta }}\left( {\Gamma _{\theta \chi }^\chi \sqrt {\frac{{{{\rm{g}}_{\chi \chi }}}}{{{{\rm{g}}_{\theta \theta }}}}} } \right)} \right],
\end{align}
with   $\Gamma_{\theta \theta}^{\chi}$ and $\Gamma_{\theta \chi}^{\chi}$ account for the Christoffel symbols given by
\begin{equation}
\Gamma_{\theta \theta}^{\chi}=-\frac{1}{2 \mathrm{g}_{\chi \chi}}\left[\frac{\partial \mathrm{g}_{\theta \theta}}{\partial \chi}\right], \quad \text{and}  \quad \Gamma_{\theta \chi}^{\chi}=\frac{1}{2 \mathrm{g}_{\chi \chi}}\left[\frac{\partial \mathrm{g}_{\chi \chi}}{\partial \theta}\right].
\end{equation}
It is clear that the time component $\mathrm{g}_{\chi \chi}$ equals zero at the points $\theta=0$ and $\theta=\pi$. This implies that the R-curvature is undefined at these locations, suggesting the presence of singularities at both points. However, it remains well-defined at all other locations within the $n$ spin-1/2 state space. Substituting $\mathrm{g}_{\theta \theta}$ and $\mathrm{g}_{\chi \chi}$ into the equation \eqref{g}, we get the relevant R-curvature as follows
\begin{equation}\label{h}
{R} = \frac{{16}}{{n}}\left[ {2 + \frac{{(2{ n} - 3){{\sin }^2}\theta  - 3({ n} - 1)}}{{{{\left( {(2{n} - 3){{\sin }^2}\theta  - 2({n} - 1)} \right)}^2}}}} \right].
 		\end{equation}
The result \eqref{h} reveals that the curvature of the $n$ spin-1/2 state space depends explicitly on the initial parameter $\theta$ and the particle number $n$, but remains entirely independent of $\chi$, which encodes the system's temporal evolution. This indicates that the geometry of the quantum state space, as captured by the curvature, is determined solely by the initial configuration and not by the dynamical parameter governing time evolution. Furthermore, the R-curvature \eqref{h} satisfies the periodic condition ${R}(\theta) = R(\theta + \pi)$. This matches our results, as the obtained space identified by the metric tensor \eqref{f} is a Bounded, two-dimensional space.

Fig.\ref{Curve}(b) depicts the behavior of the Riemannian curvature as a function of the initial parameter $\theta$ and the number of spins $n$. The plot reveals a clear geometric symmetry of the quantum state manifold with respect to the central axis at $\theta=\frac{\pi}{2}$, which corresponds to the point of minimum curvature. In the range $0 \leq \theta \leq {\pi}/{2}$, the curvature steadily decreases, indicating a concave bending of the manifold. Conversely, in the interval ${\pi}/{2} \leq \theta \leq \pi$, the curvature increases again, giving rise to a convex structure. This symmetric curvature profile suggests that the overall geometry of the quantum state space resembles a dumbbell shape, with the most compressed region occurring near maximal superposition states. Additionally, for $n>2$, the curvature becomes negative in the vicinity of $\theta={\pi}/{2}$, signaling regions of strong geometric deformation, often associated with enhanced entanglement or critical behavior. This observation is consistent with the results reported in Ref. \cite{Krynytskyi2019}, further confirming that the complexity of the state space geometry grows with the system size. To conclude this section, it is instructive to examine the intimate relationship between the geometric phase and the underlying geometry and curvature of the quantum state manifold. This geometric attribute not only reflects the global structure of the state space but also serves as a key physical resource in advanced quantum technologies \cite{Wang2001,Vedral2003}. The following section is devoted to exploring this profound connection in detail.
\section{Geometric phase obtained by the $n$ spin-$1/2$ system}\label{sec4}
 Once we have explored the geometry and topology of the $n$ spin-$1/2$ state space specified by the FS-metric \eqref{f}, {  we now turn our attention to a fundamental manifestation of quantum geometry: the geometric phase accumulated by the evolved state \eqref{c} under both general and cyclic evolutions. This phase carries profound physical and computational significance.}
 \subsection{Geometric phase for an arbitrary motion}
In this scenario, we consider that the $n$ spin$-1/2$ system moves arbitrarily along a specific trajectory within the Bounded two-dimensional space \eqref{f}. In this context, the geometric phase gathered by the evolved state \eqref{c} is represented as \cite{Oxman2011}
\begin{equation}\label{m}
{\gamma _g}(t) = \underbrace {\arg \langle {\psi _i}|\psi (t)\rangle }_{\gamma_{\operatorname{tot}}} - \underbrace {{\rm{Im}}\int {\langle \psi (t)|\frac{\partial}{\partial t} |\psi (t)\rangle } dt}_{\gamma_{\operatorname{dyn}}},
\end{equation}
where $\gamma_{\text{tot}}$ and $\gamma_{\text{dyn}}$ represent the total phase and the dynamic phase, respectively. To compute this geometric phase, we first need to evaluate the total phase acquired by the system. The overlap between the initial state \eqref{b} and the final state \eqref{c} is given by
\begin{equation}\label{n}
\langle\psi_i| \psi(t)\rangle=\sum\nolimits_{p=0}^{n}\mathrm{C}_n^p\tan^{2p}
\frac{\theta}{2}\cos^{2n}
\frac{\theta}{2}
 \exp{\left(-\frac{i\chi(n-2p)^2}{4}\right)}.
\end{equation}
It follows that the overall phase gained by the evolved state \eqref{c} is given by
\begin{small}
\begin{equation}\label{o}
 			\gamma_{\operatorname{tot}}=-\tan^{-1}\left(
 			\frac{\sum\nolimits\nolimits_{p=0}^{n}\mathrm{C}_n^p\tan^{2p}
\frac{\theta}{2}\cos^{2n}
\frac{\theta}{2}
\sin\left(\frac{\chi(n-2p)^2}{4}\right)}{\sum\nolimits\nolimits_{p=0}^{n}\mathrm{C}_n^p\tan^{2p}
\frac{\theta}{2}\cos^{2n}
\frac{\theta}{2}
\cos\left(\frac{\chi(n-2p)^2}{4}\right)}
 			\right).
\end{equation}
\end{small}
Notably, the overall phase \eqref{o} consists of two distinct components: the first is of geometric origin and is intrinsically related to the geometric and topological features defining the state spaces of these systems \cite{Oxman2011,Kolodrubetz2013}. This geometric component appears through the dependence of the overall phase \eqref{o} on both the R-curvature \eqref{h} and the FS-metric element $\mathrm{g}_{\chi \chi}$ \eqref{f}, as they are all functions of the parameters $n$ and  $\theta$. The second component is of dynamical origin, originating from the time-dependent evolution of Hamiltonian eigenstates \eqref{3}. The overall phase \eqref{o} shows a non-linear dependence on time and meets the periodic requirements
\begin{equation}
\gamma_{\operatorname{tot}}(\chi+4\pi)=\gamma_{\operatorname{tot}}(\chi)\qquad \text{for}\; n \; \text{integer}, \quad \text{while} \quad 
\qquad\gamma_{\operatorname{tot}}(\chi+8\pi)=\gamma_{\operatorname{tot}}(\chi)\qquad \text{for}\; n \; \text{half-integer}.
\end{equation}
From the equation \eqref{m}, we can determine the dynamical phase accumulated during the evolution process. Explicitly, we obtain 
\begin{equation}\label{u}
\gamma_{\operatorname{dyn}}=-n\frac{\chi}{4}\left[(n - 1) \cos^2 \theta + 1\right].
\end{equation}
which is proportional to the evolution time. This highlights that the dynamical phase indicates the duration necessary for the system to perform this evolution. Thus, the geometric phase accumulated by the $n$ spin$-1/2$ system \eqref{a} for a given path within the space of states \eqref{f} is given by
\begin{small}
\begin{align}\label{p}
\gamma_{\operatorname{g}}=-\tan^{-1}\left(
 			\frac{\sum\nolimits\nolimits_{p=0}^{n}\mathrm{C}_n^p\tan^{2p}
\frac{\theta}{2}\cos^{2n}
\frac{\theta}{2}
\sin\left(\frac{\chi(n-2p)^2}{4}\right)}{\sum\nolimits\nolimits_{p=0}^{n}\mathrm{C}_n^p\tan^{2p}
\frac{\theta}{2}\cos^{2n}
\frac{\theta}{2}
\cos\left(\frac{\chi(n-2p)^2}{4}\right)}
 			\right)+n\frac{\chi}{4}\left[(n - 1) \cos^2 \theta + 1\right].
\end{align}
\end{small}
 
It is apparent that the geometric phase \eqref{p} presents a nonlinear dependence on the evolution parameter $\chi$ (proportional to time), in contrast to the dynamical phase, which evolves linearly. This nonlinear behavior reflects the fact that the geometric phase is not a function of how fast the system evolves, but rather of the trajectory traced by the state vector in the curved quantum state space defined by the FS-metric \eqref{f}. It depends on both $(\theta, \chi)$, which describe the motion through the state manifold, and on the system parameter $n$, which governs the dimensionality and curvature of that manifold. The sensitivity of $\gamma_g$ to these parameters underlines its role as a geometric probe, capable of capturing both the topology and the local curvature of the quantum evolution path. From a practical perspective, this geometric phase offers a powerful tool for characterizing and controlling quantum dynamics in many-body spin systems. Its dependence on system size and initial conditions makes it particularly suitable for parameter estimation, trajectory classification, and the detection of geometric features that emerge during quantum evolution. More importantly, due to its intrinsic geometric nature, $\gamma_g$ is robust against certain types of noise and local perturbations, which makes it a key resource in the development of geometric and holonomic quantum gates. These gates, which rely on cyclic accumulation of geometric phase, form the building blocks of fault-tolerant quantum computation. Thus, the explicit form of \eqref{p} not only provides a quantitative description of the phase acquired during arbitrary quantum motion, but also serves as a foundation for the design and optimization of quantum information protocols, especially those exploiting global properties of state space geometry \cite{Kumar2005, Kleipler2018}. The upcoming subsection will focus on the specific case of geometric phase arising from cyclic evolutions. In this regime, where the system undergoes a closed trajectory, the geometric phase acquires a distinct topological character, reflecting the fundamental geometry of the resulting quantum state space \eqref{f}.
\subsection{Geometric phase in closed-path evolution}
In this new perspective, our attention is directed toward exploring the geometric phase emerging from the cyclic dynamics of the system. In this context, the wave function \eqref{c}  satisfies the cyclic criterion $|\psi({\tau})\rangle=\exp({i \gamma_{\operatorname{tot}}})|\psi(0)\rangle$ with ${\tau}$ denoting the time elapsed during the system's cyclical evolution. The AA-geometric phase (widely recognized as the Aharonov-Anandan phase) accrued by the system over a specified cyclic motion path is given by the following expression \cite{Aharonov1987,Pati1995}
\begin{equation}\label{r}
\gamma_{\operatorname{g}}^{\mathrm{AA}}=i \int_0^{{\tau}}\langle\tilde{\psi}(t)| \frac{\partial}{\partial t} |{\tilde{\psi}}(t)\rangle d t,
\end{equation}
where $| \tilde{\psi}(t)\rangle$ refers to the Anandan-Aharonov section mentioned  in \cite{Aharonov1987}. It is given in terms of the wave function \eqref{c} by $|\tilde{\psi}(t)\rangle=\exp({-i {\eta}(t)})|\psi(t)\rangle$, so that ${\eta}(t)$ is a specific smooth function fulfilling the requirement $\eta({\tau})-\eta(0)=\gamma_{\operatorname{tot}}$. In this regard, the AA-geometric phase \eqref{r} is given by
\begin{equation}\label{w}
\gamma_{\operatorname{g}}^{\mathrm{AA}}=\int_0^{{\tau}} d \gamma_{\operatorname{tot}}+i \int_0^{{\tau}}\langle\psi(t)| \frac{\partial}{\partial t} |{\psi}(t)\rangle dt.
\end{equation}
Substituting \eqref{o} and \eqref{u} into \eqref{w}, we find the AA-geometric phase gained by the $n$ spin-$1/2$ system \eqref{c} throughout a cyclic evolution expressed as
\begin{equation}\label{x1}
\gamma_{\operatorname{g}}^{\mathrm{AA}}= -\frac{1}{2}{n\pi}(n-1)\sin ^2 \theta.
\end{equation}
It is worth noting that the geometric phase \eqref{x1} acquired during closed-path motions is not influenced by the system's dynamics. Alternatively, it relies solely on the initial parameters $\theta$ and $n$ and thus depends on the choice of the initial state. This implies that the AA-geometric phase is impacted by the configuration of the relevant space of states instead of the path that the system follows throughout the evolution. Therefore, this cyclic phase proves ineffective in parameterizing and differentiating the cyclic motion trajectories followed by the considered system. 
\begin{figure}[htbp]
\begin{center}
\includegraphics[scale=0.4]{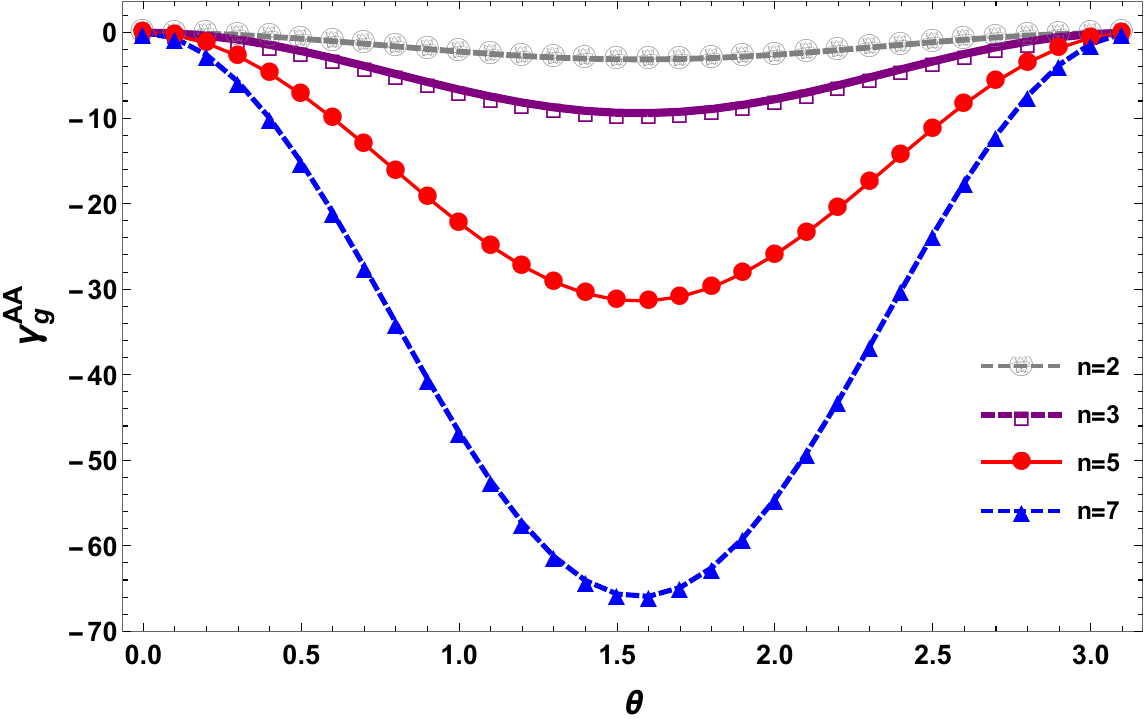}
\caption{The cyclic geometric phase \eqref{x1} against the variable $\theta$ for specific spin$-1/2$ numbers.}\label{av}
\end{center}
\end{figure}

Figure \eqref{av} illustrates the response of geometrical phase \eqref{x1} to the starting parameters $(n,\theta)$, where it displays behavior akin to the R-curvature, as seen in Figure 1. Indeed, Fig. \eqref{av} shows that cyclic geometric phase \eqref{x1} exhibits a symmetric behavior around $\theta=\pi/2$, forming a bell-shaped curve, akin to the R-curvature (see Fig. \ref{Curve}(b)). In the interval [$0, \pi /2$], the geometric phase decreases as $\theta$ increases, consistent with a concave geometry where the curvature drops. Conversely, for $\theta \in[\pi/2, \pi]$, the phase rises, reflecting a convex geometric region where the curvature increases. This symmetry about $\theta=\pi / 2$ highlights the influence of the underlying dumbbell-like structure of the quantum state manifold and reinforces the interpretation that this central point corresponds to maximal quantum superposition. As the number of spins $n$ increases, the amplitude of the phase becomes more pronounced, emphasizing the growing role of many-body entanglement in shaping the system's geometric evolution. Thus, the phase not only encodes dynamical information but also serves as a probe of the global geometry of the quantum state space. The topological phase may likewise be addressed in this context; it constitutes the part of such a geometric phase that is not influenced by any dynamic factors. It is acquired as

 \begin{equation}\label{a1}
 \gamma_{\operatorname{top}}^{\mathrm{AA}} =-\frac{\pi}{2}n^2.
 \end{equation}
 It is clear to see that the obtained topological phase \eqref{a1} is proportional to the square of the number of particles. Notably, this phase assumes fractional values for $n$ odd and multiples of $\pi$ for $n$ even. This demonstrates the significance of the topological phase in controlling the topology associated with the relevant state space, provided in equation \eqref{f}. This allows for the parametrization of the closed evolution paths followed by the system utilizing the acquired topological phase \eqref{a1}. This finding may be relevant to quantum computation, especially in the development of effective quantum circuits \cite{Johansson2012,Roushan2014}.\par 
 
 Having investigated both the cyclic and noncyclic geometric phases, we now turn to a complementary aspect of the $n$-spin-${1}/{2}$ system, its dynamical properties, which govern the evolution within the quantum state space \eqref{f}. To complete our geometric characterization, it is essential to analyze the system’s temporal evolution. In the next section, we employ the FS-metric to perform a detailed analysis of the system’s dynamics, evaluating the evolution speed and the corresponding FS-distance that separates any two quantum states within the relevant manifold. Building upon these results, we further address the quantum brachistochrone problem, which concerns determining the minimal time required for optimal evolutions.
\section{Speed and time-optimum motion for $n$ interacting spin-1/2}\label{sec3}
We shall now employ the FS-metric tensor \eqref{f}, specifying the pertinent state space \eqref{f}, to determine and analyze certain dynamic characteristics of the system. More precisely, we investigate the evolution speed and the geodesic length quantified by the FS-metric \eqref{f}. In this scope, we attempt to solve the problem related to the quantum brachistochrone \cite{Ammghar2020,Mostafazadeh2007}. Addressing this problem entails determining the time-optimal motion by maximizing the evolution rate and pinpointing the shortest evolution route that the system takes between the initial state \eqref{b} and the final state \eqref{c}. Simply put, resolving the brachistochrone problem refers to evaluating the least time required to complete a given evolution.\par

{\bf Speed of quantum motion:} To assess the evolution speed, we consider that the $n$ spin$-1/2$ system depends solely on time, with all other parameters held constant. In this perspective, the FS-metric tensor \eqref{f} rewrites
 \begin{equation}\label{a4}
 	dS^2=\frac{1}{8} n(n-1) \sin ^2 \theta\left[1+\left(2n-3\right) \cos ^2 \theta\right]d\chi^2,
 		\end{equation}
and thus the space of states \eqref{f} of the system reduces to a circle of radius $\sqrt{\mathrm{g}_{\chi \chi}}$. On the other hand, the evolution rate of a such quantum system \eqref{c} is given by \cite{Anandan1990}
 \begin{equation}\label{a5}
 V=\frac{d S}{d t}.
 \end{equation}
 Reporting the finding \eqref{a4} within \eqref{a5}, the evolution speed of the system \eqref{c} is obtained as
 \begin{equation}\label{a7}
  V=\frac{{J}}{2}\sqrt {n(n - 1){{\sin }^2}\theta \left[ {\frac{1}{2} + \left( {n - \frac{3}{2}} \right){{\cos }^2}\theta } \right]}.
 \end{equation}
Remark that the evolution rate \eqref{a7} is dependent on both the coupling constant $J$ and the choice of initial state, specified by the parameters: $\theta$ and $n$. Added to that, the higher the number of particles and the coupling constant, the faster the system moves, except for the singular points: $\theta=0$ and $\theta=\pi$, where the evolution speed \eqref{a7} vanishes $(V =0)$, irrespective of all other physical parameters. This is supported by the fact that the $n$ spin$-1/2$ state \eqref{c} and the R-curvature \eqref{g} are not defined at these points.\par
\begin{figure}[htbp]
\begin{center}
\includegraphics[scale=0.4]{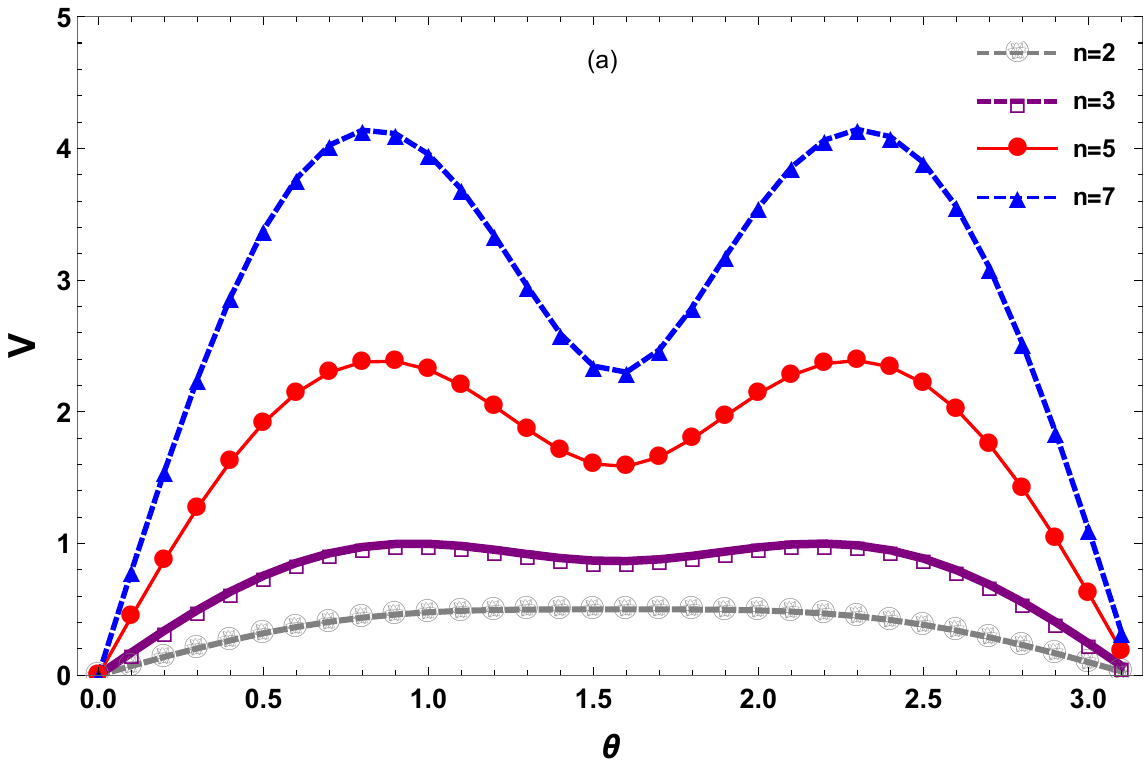}\includegraphics[scale=0.4]{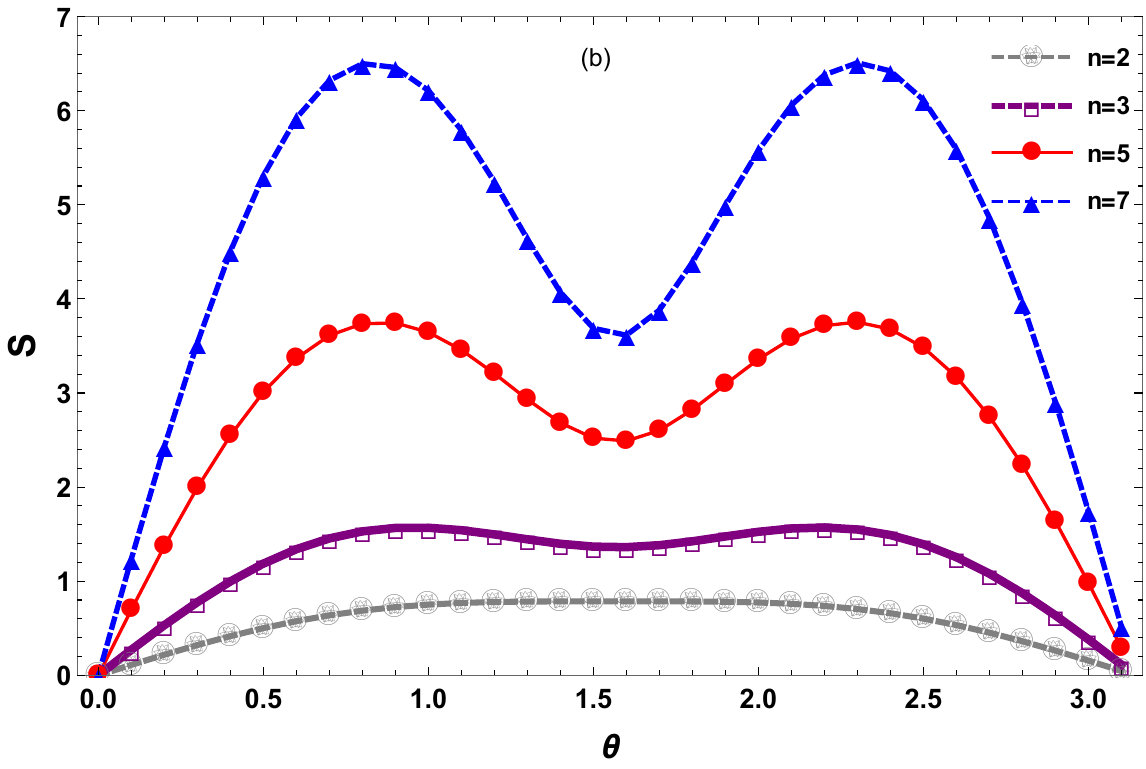}
\caption{The evolution rate (speed) [Eq. \eqref{a7}, panel (a)] and the FS-distance [Eq. \eqref{a8}, panel (b)] versus the parameter $\theta$ for certain spin$-1/2$ numbers by setting $J=1$.}\label{al}
\end{center}
\end{figure}

On the flip side, the dependence of speed on $\theta$ reveals that it is also influenced by both the R-curvature \eqref{h} and geometric phases \eqref{p} and \eqref{x1}. The pattern of the speed \eqref{a7} in relation to the initial parameters $\theta$ and $n$ is clearly depicted in figure \ref{al}(a).

{\bf Time-optimal motion:} To resolve the quantum brachistochrone problem and achieve the time-optimal motion of the considered system, we must determine the minimal time required to achieve such motion \eqref{c}. To do this, we start by assessing the maximal speed through the resolution of the equation $dV/d\theta=0$, which leads to
 		\begin{equation}
 			\left[n-2-(2n-3)\cos^2\theta\right]\sin 2\theta=0,
 		\end{equation}
 		which implies that
 		\begin{equation}
 		  \cos^2\theta_{\max}=
 		     {\frac{n-2}{2n-3}}.
 		\end{equation}
Thereby, the maximum speed attainable by the $n$ spin-$1/2$ system is given by
 		\begin{equation}
 			{V}_{\max}=\frac{1}{2}J(n-1)\sqrt{\frac{n(n-1)}{2(2n-3)}},
 		\end{equation}
 	Thus, we succeed in finding the maximal speed, which depends only on the number of particles in the system. In addition, the FS-distance that the system covers between the initial state \eqref{b} and the final state \eqref{c} can also be formulated, using the result \eqref{a5}, by
 		\begin{equation}\label{a8}
{S}=\frac{{\chi}}{2}\sqrt {n(n - 1){{\sin }^2}\theta \left[ {\frac{1}{2} + \left( {n - \frac{3}{2}} \right){{\cos }^2}\theta } \right]}.
 \end{equation}

From a geometric standpoint, since the speed is proportional to the Fubini-Study line element $dS/d t$, this plot captures how fast the quantum state moves along its trajectory on the curved, two-dimensional manifold defined by $(\theta, \chi)$. The symmetry of the curves about $\theta=\pi / 2$ and the periodic boundary behavior at $\theta=0$ and $\theta=\pi$ are consistent with the periodic curvature $R(\theta)=R(\theta+\pi)$, confirming the spherical topology of the state space (see Fig. \ref{Curve}(a)). Physically, the speed reaches its maxima near  $\theta \approx \pi/3$ and $2\pi/3$, corresponding to the equatorial region of the manifold, where the curvature is minimal and the quantum state is most sensitive to collective spin interactions. In contrast, at the poles $(\theta=0, \pi)$, where the curvature becomes large and positive, the motion slows down significantly, indicating regions of dynamical freezing. The growth in the peak speed with increasing $n$ highlights how larger spin systems, through stronger entanglement and interaction effects, allow faster traversal of the state space while preserving its underlying topology. In other words, the curvature and topology of the quantum state manifold \eqref{f} directly influence the evolution speed: flatter regions near the equator accelerate motion, while highly curved polar regions suppress it. This identifies curvature as a key geometric resource for optimizing quantum control. These findings emphasize the deep link between curvature and dynamics, providing a geometric lens through which quantum control can be optimized by targeting regions of the state manifold where evolution inherently accelerates \cite{p2}.\par

Notice that the evolution speed \eqref{a7} is time-independent, this implies that the FS-distance \eqref{a8} exhibits a linear behavior with time. {From Figs. (\ref{al}a) and (\ref{al}b), it is evident that the FS-distance \eqref{a8} and the corresponding evolution speed exhibit remarkably similar behavior across the state space \eqref{f}. This strong correlation implies that the FS-distance, like the speed, is intrinsically governed by the underlying curvature and topology of the quantum state manifold. Specifically, the FS-distance attains its maximum in the flatter equatorial regions of the manifold, where the Riemann curvature is minimal and the state space locally resembles a Euclidean geometry, enabling the quantum state to traverse longer distances over time. In contrast, near the highly curved polar regions, the FS-distance shrinks significantly, reflecting the geometric contraction induced by positive curvature that limits the extent of accessible state evolution. This geometric modulation is a direct consequence of the manifold's dumbbell-like shape, rooted in its spherical topology. Thus, the FS-distance not only quantifies the distinguishability of quantum states but also encodes rich geometric information about how curvature and topology constrain or facilitate the system's evolution. This geometric sensitivity makes the FS-distance a powerful diagnostic tool in optimizing quantum protocols that rely on efficient and controllable state transitions \cite{p3,p4}.} From this evaluation, the FS-distance \eqref{a8} exhibits a local minimum when $\theta=\pi/2$, expressed as
 	\begin{equation}\label{a9}
  {S}_{\min}=n\frac{{\chi}}{2}\sqrt{\frac{1}{2}\left(1-\frac{1}{n}\right)}.
 \end{equation}	
 Hence, the shortest period of time needed to perform any time-optimum motion, of the considered system, reads
 \begin{equation}\label{a10}
 {t}_{\min}=\frac{{S}_{\min}}{V_{\max}}=\frac{\sqrt{t^2(2n-3)}}{(n-1)}, \quad\text{with}\quad n>1.
 \end{equation}
 This result represents the minimal time needed to traverse the shortest geodesic on the manifold of quantum states \eqref{a4}, effectively combining the maximum attainable speed with the minimal FS-distance. It encapsulates the geometric constraint for achieving optimal evolution: the system must evolve along the most direct path between two quantum states with the greatest possible velocity. Therefore, the array of quantum states that can achieve optimal evolution is derived from the transformation
 \begin{equation}
 		\left|\psi_i\right\rangle \rightarrow|\psi({t}_{\min})\rangle=\exp({-i {H} {t}_{\min}})\left|\psi_i\right\rangle,
 		\end{equation}
and the corresponding optimal state space is identified by the metric tensor 
 		\begin{small}
 	\begin{equation}\label{a11}
 	dS^2_{\mathrm{op}}=\frac{1}{8} n(n-1) \sin ^2 \theta\left[1+\left(2n-3\right) \cos ^2 \theta\right]d\chi^2_{\min},
 		\end{equation}	
 		\end{small}
  with $\chi_{\min }=J t_{\min }$. This optimal state manifold retains the spherical topology of the original quantum space but is geometrically reshaped by the interaction strength and spin count. Importantly, the minimal time $t_{ {\min }}$ is directly affected by the number of particles $n$, highlighting the profound influence of topology and curvature on quantum evolution. For two-spin systems ($n=2$), the optimal and standard durations coincide ($t_{\min }=t$), reflecting a relatively flat and less curved geometry. However, when $n\geq3$, the optimal time is strictly shorter than the standard evolution time, indicating that the geometric structure allows for faster trajectories as the curvature flattens near the equator and the topological complexity increases. In the thermodynamic limit $(n \rightarrow \infty)$, $t_{{\min }}$ tends to zero, and the optimal manifold effectively straightens, as its local radius $\sqrt{g_{\chi_{\min } \chi_{\min }}}$ grows without bound, implying a transition from a curved to a nearly flat geometry.\par
 From a practical standpoint, this dependence of $t_{\text {min }}$ on both the interaction topology and the spin count has far-reaching implications. In quantum information processing, minimizing operation time is essential for reducing decoherence and maximizing fidelity \cite{p6}. These results demonstrate that entanglement and many-body effects are not merely complications but can be harnessed to accelerate quantum evolution. Furthermore, because the optimal manifold geometry reflects the system's symmetry and interaction structure, sudden changes in $t_{{\min}}$ or its associated curvature may serve as indicators of quantum phase transitions. In this respect, the evolution time $t$ and particle number $n$ function not only as dynamical variables but also as geometric probes revealing the deeper topological and physical nature of the quantum system.\par 
  At this point, we have explored the key geometric and dynamical characteristics of the underlying quantum state manifold through the FS-metric, associated geometric phases, evolution speed, and FS-distance. A natural next step is to uncover how these properties are shaped by quantum entanglement. As a fundamental quantum resource, entanglement not only reshapes the geometry of the state space but also exerts a profound influence on the system’s dynamical behavior. In what follows, we unveil this intricate interplay, showing how entanglement influences the accumulation of geometric phase, modulates evolution speed, alters the distance between quantum states, and reduces the minimal time required for optimal evolutions.
\section{Geometric and dynamic aspects of the entanglement amount exchanged between two spins}\label{sec5}
In this final section, we limit the above system to two spin-1/2 particles with the Ising-type interaction. The goal is to examine the quantum entanglement exchanged between these two particles from two viewpoints: the first one is geometric and deals with the study of the entanglement influence over the geometric structures established above, such as the FS-metric, R-curvature, and the cyclic and non-cyclic geometric phases. The second one is dynamic and explores the impact of entanglement on both the evolution speed and the FS-distance achieved by the system. Moreover, we use the degree of entanglement to address the problem related to the quantum Brachistochrone.
\subsection{Geometric quantification of entanglement in the two spin-1/2 system} 	
From this vantage point, the evolving state that defines the system of two spin-1/2 particles is expressed as
 \begin{align}\label{a12p}
 		|\psi (t)\rangle  = \exp ( - i\chi (t)){\cos ^2}\frac{\theta }{2}\left| {\frac{1}{2},\frac{1}{2}} \right\rangle  + \frac{1}{2}\exp (i\varphi )\sin \theta \left( {\left| {\frac{1}{2}, - \frac{1}{2}} \right\rangle  + \left| { - \frac{1}{2},\frac{1}{2}} \right\rangle } \right) + \exp (i(2\varphi  - \chi (t))){\sin ^2}\frac{\theta }{2}| - \frac{1}{2}, - \frac{1}{2}\rangle,
 		\end{align} 	
and consequently, the space of two spin-$1/2$ states where the system motion unfolds is characterized by the ensuing FS-metric
 \begin{equation}\label{a14}
 d S^2=\frac{1}{2} d \theta^2+\frac{1}{4} \left(1-\cos^4 \theta\right) d \chi^2.
\end{equation}
 To examine the amount of quantum correlations present in the two-spin system \eqref{a12p}, we use the geometric measure of entanglement as an indicator  \cite{k1,k2}. It is defined in the literature by \cite{Frydryszak2017}
\begin{equation}\label{a12}
\mathtt{E}=\frac{1}{2}\left(1-\left|\left\langle\boldsymbol{\sigma}_i\right\rangle\right|\right),
\end{equation}
where the average value of one of two spins $(i=1,2)$ can be developed in the Cartesian basis as 
\begin{equation}
\left\langle\boldsymbol{\sigma}_i\right\rangle=\left\langle\sigma_i^x\right\rangle \mathbf{i}+\left\langle\sigma_i^y\right\rangle \mathbf{j}+\left\langle\sigma_i^z\right\rangle \mathbf{k}
\end{equation}
Then, evaluating the average value of $i$-$th$ spin within the evolved state \eqref{a12p} and putting it in \eqref{a12}, we give the geometric measure of entanglement under form 
\begin{equation}\label{a13}
\mathtt{E}=\frac{1}{2}\left[1-\sqrt{1-\sin^4 \theta \sin^2 \chi}\right].
\end{equation}
From this result \eqref{a13}, we observe that the entanglement of the considered system is given in terms of its dynamic degrees of freedom $(\chi, \theta)$, showing that the entanglement evolution depends on the choice of the evolution path pursued by the system within the two-spin state space \eqref{a14}. In addition, for a specific value of $\theta$, excluding $\theta=0$ or $\pi$, the entanglement undergoes periodic variations over time. We attain the states of highest level of entanglement for $\chi=\pi/2 \text{ or } 3\pi/2$, whilst the disentangled states $(\mathtt{E}=0)$ are attained for $\chi=0 \text{ or } \pi$. More precisely, in the region $\chi\in[0,2\pi]$, the entanglement of evolving state \eqref{a12p}  attains two maxima $\mathtt{E}_{\max}=1/2[1-\sqrt{1-\sin^4 \theta}]$ at the positions $\chi=\pi/2, 3\pi/2$ , and two minimums $\mathtt{E}=0$ at the positions $\chi=0, \pi$. { This is clearly depicted in Fig. \eqref{amk23}, illustrating the behavior of the entanglement against the dynamical parameter $\chi$. Indeed,
\begin{figure}[htbp]
\begin{center}
\includegraphics[scale=0.4]{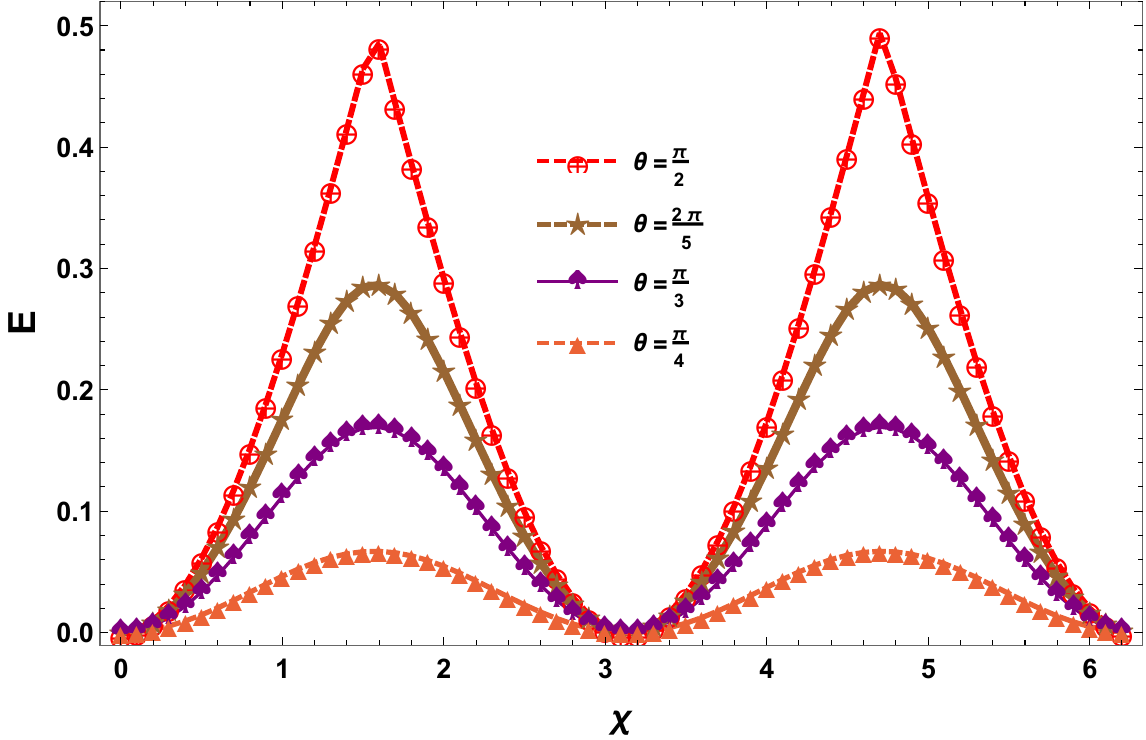}
\caption{The geometric measure of entanglement \eqref{a13} versus the dynamic parameter $\chi$  for some values of $\theta$.}\label{amk23}
\end{center}
\end{figure}
we observe that the entanglement shared between the two spins exhibits a periodic dependence on the evolution parameter $\chi$, revealing the intrinsically cyclic nature of the system's dynamics. Notably, the amplitude of this entanglement is strongly modulated by the initial state parameter $\theta$, with the highest entanglement achieved for intermediate values such as $\theta=\frac{\pi}{2}$. In contrast, for extremal values $\theta=0$ or $\pi$, the entanglement remains zero throughout the evolution ($\mathrm{E}=0$ for all $\chi$). This behavior can be directly traced back to the choice of initial quantum states: in these limiting cases, the system begins in the product states $\left|\psi_i\right\rangle=\left|\frac{1}{2}, \frac{1}{2}\right\rangle$ or $\left|-\frac{1}{2},-\frac{1}{2}\right\rangle$, which are eigenstates of the system's Hamiltonian. Since these states are stationary under unitary evolution generated by the Hamiltonian, the system undergoes no nontrivial trajectory in state space and therefore fails to generate entanglement. This underlines a key point: entanglement generation is not merely a consequence of interaction, but requires the system to traverse a nontrivial path through its curved state space, with the geometry and initial conditions jointly determining the extent to which entanglement can arise.\par In the particular scenario where  $\theta=\pi/2$,  the expression for the geometric measure of entanglement simplifies considerably and takes the form
\begin{equation}
\mathtt{E} = \frac{1}{2}\left[ {1 - \left| {\cos \chi } \right|} \right],
\end{equation}
This relation reveals a clear periodic behavior, with entanglement reaching its maximum value of $\mathrm{E}_{\max }={1}/{2}$ at specific points along the evolution, namely $\chi={\pi}/{2}$ and $\chi={3\pi}/{2}$. These particular instants correspond to configurations where the quantum state becomes maximally entangled, a feature characteristic of Schrödinger cat-like superpositions \cite{Molmer1999,Leibfried2005}. these states are critical in quantum information protocols, as they encapsulate strong quantum correlations that are essential for tasks such as quantum metrology and fault-tolerant quantum computing \cite{c5,c6,c06}.}
\subsection{Geometric facet of quantum entanglement} 		
To better understand the geometric manifestation of the quantum correlations between the two particles under analysis, we will provide a thorough explanation focusing on the interconnections between entanglement and the geometric attributes investigated above.  In fact, by including the finding \eqref{a13} in \eqref{a14}, we obtain the FS-metric specifying the two-spin space \eqref{a14} in relation with the entanglement level of the system
\begin{align}\label{a15}
d{S^2} =& \frac{{{{(2\mathtt{E} - 1)}^2}d{\mathtt{E}^2}}}{{16(\mathtt{E}{{(1 - \mathtt{E})})^{3/2}}\left( {\left| {\sin \chi } \right| - 2\sqrt {\mathtt{E}(1 - \mathtt{E})} } \right)}} + \frac{{(2\mathtt{E} - 1)d\mathtt{E}d\chi }}{{4\sqrt {\mathtt{E}(1 - \mathtt{E})} \left( {\left| {\sin \chi } \right| - 2\sqrt {\mathtt{E}(1 - \mathtt{E})} } \right)\tan \chi }}\notag\\[5px]&+\frac{\sqrt {\mathtt{E}(1 - \mathtt{E})} }{\sin^2\chi} \left[ {\frac{\cos^2\chi}{{4\left( {\left| {\sin \chi } \right| - 2\sqrt {\mathtt{E}(1 - \mathtt{E})} } \right) }} + \left( {\left| {\sin \chi } \right| - \sqrt {\mathtt{E}(1 - \mathtt{E})} } \right)} \right]d{\chi ^2},
\end{align}
which can be articulated in the following diagonal form
\begin{equation}\label{bra}
d{S^2} = \frac{1}{{8{\mathtt{E}_r}\left( {1 - {\mathtt{E}_r}} \right)}}d\mathtt{E}_r^2 + \frac{1}{4}{\mathtt{E}_r}\left( {2 - {\mathtt{E}_r}} \right)d{\chi ^2},
\end{equation}
{ where the reduced entanglement $\mathrm{E}_r=\sqrt{4 \mathrm{E}(1-\mathrm{E}) / \sin ^2 \chi}$, ranging in the interval $[0,1]$, emerges as a fundamental geometric descriptor. This formulation \eqref{bra} has both theoretical and practical significance. On the theoretical side, it provides a bridge between quantum information measures (like entanglement) and differential geometric tools (such as Riemannian metrics and curvature). Practically, since both $\mathtt{E}$ and $\chi$ are, in principle, experimentally accessible, the entanglement-induced geometry of quantum state space becomes observable and testable in controlled systems, e.g., trapped ions, superconducting qubits, or cold atoms \cite{c8,c9,c10}.} Significantly, quantum entanglement contributes to the reduction of the state space dimensions. More precisely, the two-spin-1/2 states possessing the same entanglement level  $(\mathtt{E}=\text{constant})$ reside on within one-dimensional enclosed spaces defined by
\begin{equation}
d{S^2} = \frac{\sqrt {\mathtt{E}(1 - \mathtt{E})} }{\sin^2\chi} \left[ {\frac{\cos^2\chi}{{4\left( {\left| {\sin \chi } \right| - 2\sqrt {\mathtt{E}(1 - \mathtt{E})} } \right) }} + \left( {\left| {\sin \chi } \right| - \sqrt {\mathtt{E}(1 - \mathtt{E})} } \right)} \right]d{\chi ^2}.
\end{equation}
They form closed loops along the metric component $\mathrm{g}_{\chi\chi}$, over the entire state space \eqref{a15}. The quantum states exhibiting an identical level of reduced entanglement $(\mathtt{E}_r=\text{constant})$ are found within the circles specified by
\begin{equation}
 dS^2=\frac{1}{4}{\mathtt{E}_r}\left( {2 - {\mathtt{E}_r}} \right)d{\chi ^2},
\end{equation}
featuring radii $R=(1/2)\sqrt{ \mathtt{E}_r\left(2-\mathtt{E}_r\right)}$, reliant on the specific degree of reduced entanglement. { This geometric confinement elucidates the role of entanglement in reducing the effective dimensionality of the quantum state space \eqref{a15}. In essence, quantum correlations impose constraints on the system's accessible configurations (the accessible degrees of freedom), thereby inducing curvature in the underlying manifold and altering the geodesic structure of quantum evolution. This interpretation is consistent with the broader understanding that entanglement not only encodes nonclassical correlations but also shapes the geometry and topology of quantum state spaces.}\par
\begin{figure}[htbp]
\begin{center}
\includegraphics[scale=0.3]{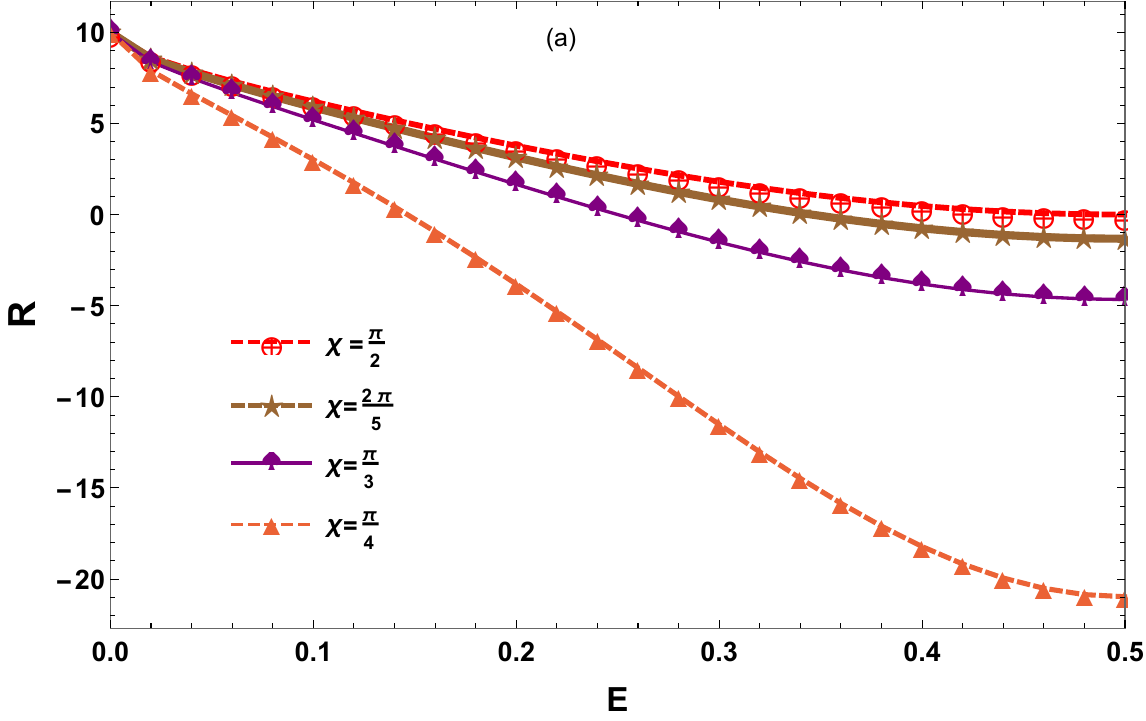}\includegraphics[scale=0.25]{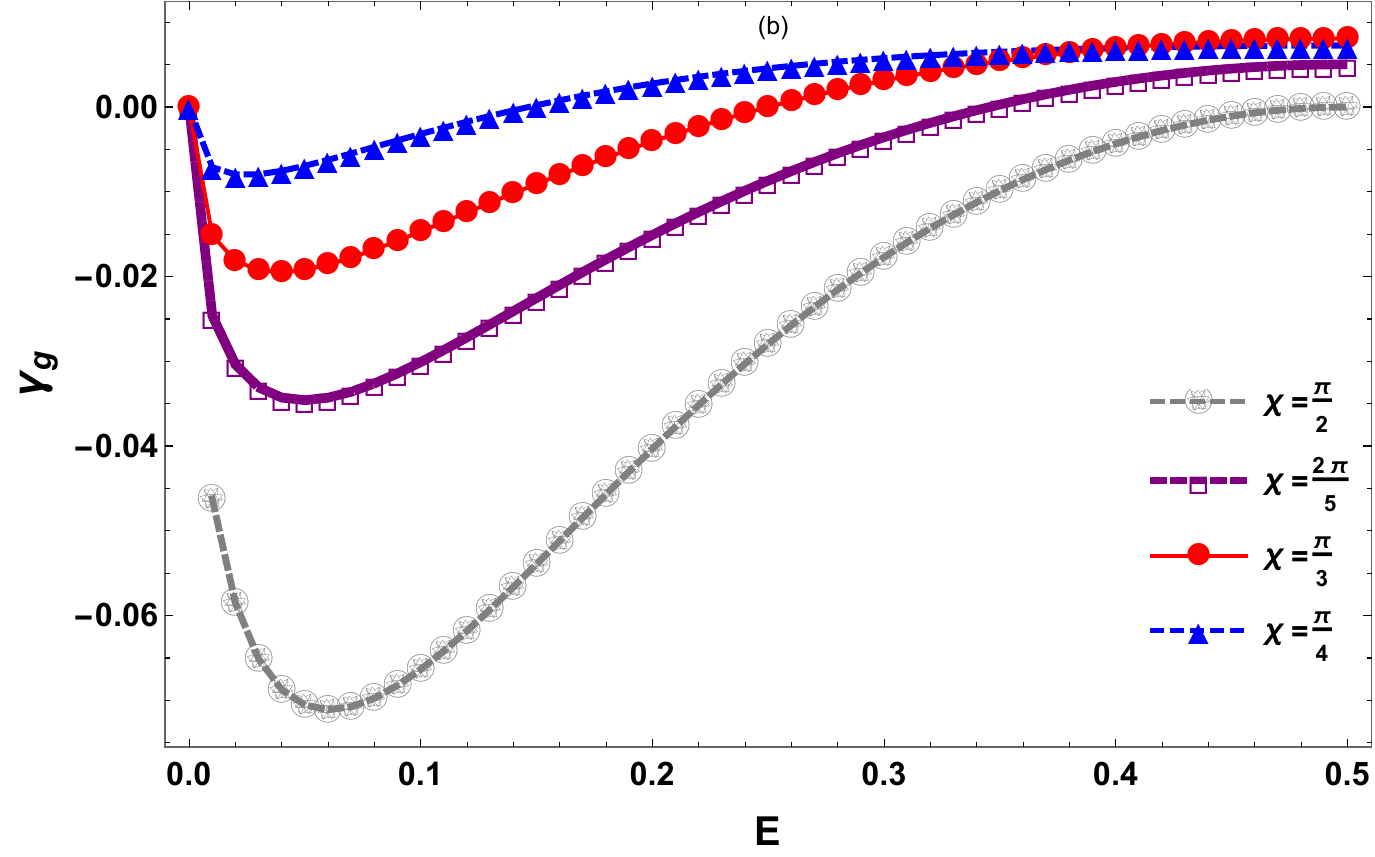}\includegraphics[scale=0.3]{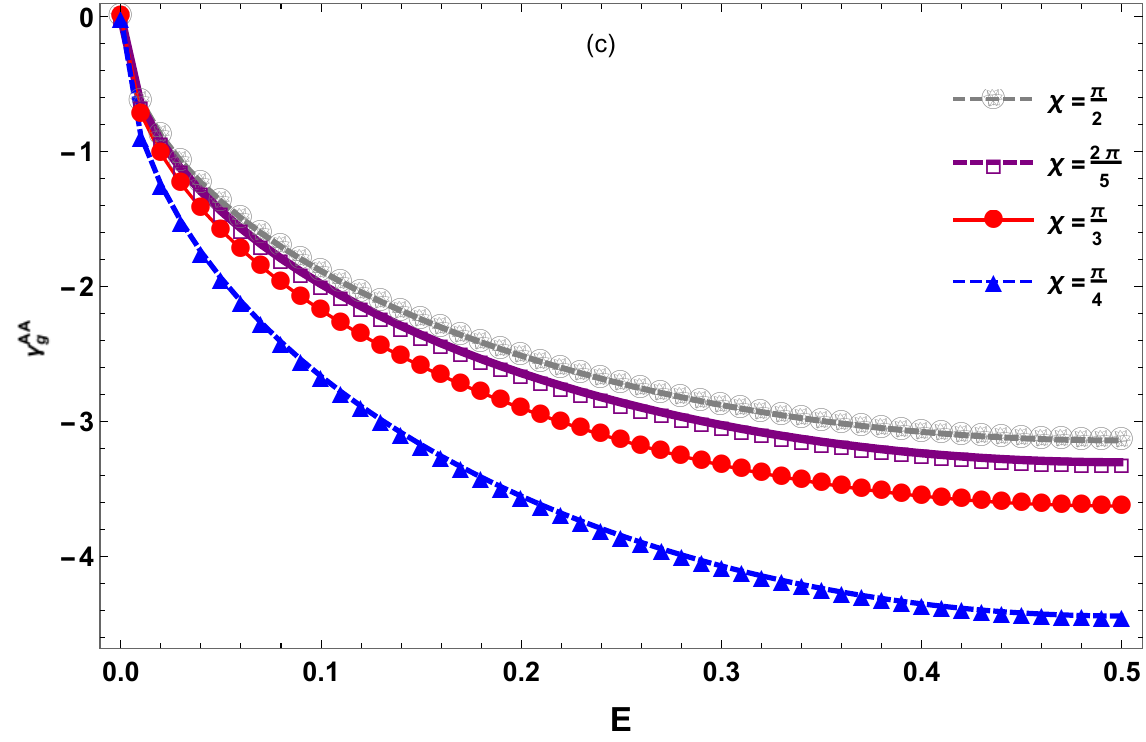}
\caption{The R-curvature [Eq. \eqref{a16}, panel (a)], geometric phase [Eq. \eqref{a19}, panel (b)], and AA-geometric phase [Eq. \eqref{a20}, panel (c)] versus the geometric measure of entanglement \eqref{a13} for some values of $\chi$ with $\theta=\pi/2$ ($\mathtt{E}\in[0,\mathtt{E}_{\max}=0.5]$).}\label{amk}
\end{center}
\end{figure}

 Within the same framework, we investigate the impact of entanglement on the R-curvature across the space of two-spin states \eqref{a15}. Introducing the finding \eqref{a13} into \eqref{h}, we establish the state space curvature in relation to the entanglement level of the system 
\begin{equation}\label{a16}
R = 8\left( {2 + \frac{{ {2\sqrt {\mathtt{E}(1 - \mathtt{E})}  - 3\left| {\sin \chi } \right|} }}{{4{{\left( {\sqrt {\mathtt{E}(1 - \mathtt{E})}  - \left| {\sin \chi } \right|} \right)}^2}}}\left| {\sin \chi } \right|} \right).
\end{equation}
This outcome underscores, yet again, the noticeable dependence of the geometry of state space on the entanglement amount conveyed between the two particles. In the case where $\chi=0$  (the static condition), the R-curvature does not depend on $\mathtt{E}$ (entanglement), and assumes a constant value of $(R=16)$, aligning with the curvature of the initial state sphere \eqref{a17}. Conversely, when $\chi>0$, indicating a dynamic case, it becomes dependent on $\mathtt{E}$. To illustrate further, we present the behavior relative to entanglement in figure \ref{amk}(a). We see that as the entanglement between the two spins increases, the R-curvature of the state space correspondingly decreases for all values of $\chi$. This monotonic decline reflects a fundamental geometric effect of quantum correlations: increasing entanglement reshapes the manifold from a highly curved (spherical-like) geometry into a progressively flatter or negatively curved (hyperbolic-like) space. In particular, when $\chi$ becomes sufficiently large, the curvature not only decreases more sharply but also turns negative beyond a critical entanglement threshold, as defined by the inequality
\begin{equation}\label{a17kkk}
\left({ {2\sqrt {\mathtt{E}(1 - \mathtt{E})}  - 3\left| {\sin \chi } \right|} }\right)\left| {\sin \chi } \right|<{-8{{\left( {\sqrt {\mathtt{E}(1 - \mathtt{E})}  - \left| {\sin \chi } \right|} \right)}^2}}.
\end{equation}
This sheds light on the critical role of quantum correlations within the two-spin system in broadening or compactifying the space of physical states \eqref{a15}. Remarkably, the non-entangled states $(\mathtt{E}=0)$ occupy regions of greatest curvature $R_{\max}=10$, whereas those with the maximal degree of entanglement $\mathtt{E}_{\max}=1/2[1-\sqrt{1-\sin^4 \theta}]$ are found in regions of minimal curvature.
\begin{equation}\label{a18}
{{R}_{\min }} = 8\left[ {2 + \frac{{{{\sin }^2}\theta  - 3\left| {\sin \chi } \right|}}{{{{\left( {{{\sin }^2}\theta  - 2\left| {\sin \chi } \right|} \right)}^2}}}\left| {\sin \chi } \right|} \right].
\end{equation}
{ 
 These findings emphasize the pivotal role of entanglement in shaping the topology and geometry of the accessible quantum state space \eqref{a15}. Consequently, knowing the degree of entanglement in the system \eqref{a12} allows us to infer its position and behavior within this curved manifold, reinforcing the deterministic nature of geometric quantum mechanics. This approach, rooted in the differential geometry of the resulting state space, highlights how entanglement shapes the structure of quantum phase space and guides the system's evolution along geodesic trajectories \cite{Brody2001,Zhang1995}.}\par
The behavior of the geometric phase relative to the entanglement warrants discussion here. Indeed, introducing the equation \eqref{a13} into \eqref{p}, we give the geometric phase, accumulated by the two-spin state \eqref{a12p},  in terms of the geometric measure of entanglement as follows
\begin{equation}\label{a19}
 {\gamma_{{\rm g}} } =  - \arctan \left[ {\frac{{\left( {\left| {\sin \chi } \right| - \sqrt {\mathtt{E}(1 - \mathtt{E})} } \right)\sin \chi }}{{\left( {\left| {\sin \chi } \right| - \sqrt {\mathtt{E}(1 - \mathtt{E})} } \right)\cos \chi  + \sqrt {\mathtt{E}(1 - \mathtt{E})} }}} \right] + \chi \left( {\begin{array}{*{20}{c}}
{1 - \sqrt {\frac{{\mathtt{E}(1 - \mathtt{E})}}{{{{\sin }^2}\chi }}} }
\end{array}} \right).
  \end{equation}
In light of this result, we observe that the geometric phase is characterized by two crucial variables: entanglement and time. This suggests that it relies on each point, specifying a physical state of the system over the state space \eqref{a15}.  As a result of this, we deduce that the geometric phase is influenced by both the geometry of the state manifold and the motion path followed by the system. In addition, the geometric phase's reliance on two measurable physical parameters, $\mathtt{E}$ and $\chi$, facilitates its experimental assessment during any system evolution process. This finding holds great importance since it can be utilized in the realization of quantum logic gates founded on geometric phases. To further clarify the connection between the geometric phase and entanglement, we have illustrated, in Figure \ref{amk}(b), the dependence of Eq. \eqref{a19} vis-a-vis the entanglement for various values of $\chi$ with $\theta$ set to $\pi/2$. We find that the geometric phase \eqref{a19} achieved by the evolved state \eqref{a12p}, transitioning from the disentangled state $(\mathtt{E}=0)$ to the state with the greatest degree of entanglement $(\mathtt{E}_{\max}=0.5)$, shows an approximate parabolic trend. Thus, we can break down its evolution into two main parts: the first part pertains to the decline of the geometric phase across the entanglement range $\mathtt{E}\in[0,\mathtt{E}{\text{c}}]$, where $\mathtt{E}{\text{c}}$ denotes the critical entanglement point, at which the geometric phase achieves a minimum value (see figure \ref{amk}(b)). It is given analytically as
\begin{small}
\begin{equation}
 \mathtt{E}_{\text{c}}=\frac{1}{2}\left[ {1 - \sqrt {2 + 2\cos \chi  + \cos 2\chi  - {{2\left( {1 + \cos \chi } \right)\operatorname{sinc} \chi }}{} + \frac{{8{{\cos }^8}\frac{\chi }{2}{{\sin }^4}\frac{\chi }{2}\sqrt {{\chi ^3}\left( {2 - \chi \cot \frac{\chi }{2}} \right){{\csc }^5}\frac{\chi }{2}{{\sec }^{11}}\frac{\chi }{2}} }}{{{\chi ^2}}}} } \right].
\end{equation}
\end{small}
In this part, the wave function \eqref{a12p}  gains a geometric phase of negative sign, which can be considered as the part of this phase that the system has relinquished. In geometric terms, we can state that throughout parallel transport, the quantum state vector \eqref{a12p} rotates clockwise, forming an angle with a negative sign concerning the non-entangled state, which is the beginning state. Therefore, we conclude that, within the range  $[0,\mathtt{E}_{\text{c}}]$, the quantum entanglement among the two particles results in a decrease in geometric phase. The second part refers to the increase in this phase along the range $\mathtt{E}\in [\mathtt{E}_{\text{c}},0.5]$, demonstrating reverse behavior, the wave function \eqref{a12p} collects a positive geometric phase of positive sign, which can be regarded as the geometric phase portion obtained by the system. From a geometric perspective, we can state that while undergoing parallel transport, the state vector \eqref{a12p} turns counterclockwise, creating a positive angle with respect to the non-entangled state. Consequently, we deduce that in the range $[\mathtt{E}_{\text{c}},0.5]$, the quantum correlations promote the accumulation of the geometric phase.\par

As a result, the behavior of the geometric phase in relation to entanglement exhibits approximate symmetry around the critical value $\mathtt{E}_{\text{c}}$, which is largely attributable to the dumbbell structure associated with the relevant space of states \eqref{a15}. From a hands-on standpoint, we deduce that quantum entanglement can be utilized empirically to manipulate and manage the geometric phase resulting from the evolution of such quantum systems.\par
The cyclic geometric phase deserves to be investigated through the lens of entanglement. In effect, reporting the result \eqref{a13} into \eqref{x1}, we give the AA-geometric phase accumulated by the evolved state \eqref{a12p} in relation to the geometric measure of entanglement as
\begin{equation}\label{a20}
\gamma_{\operatorname{g}}^{\mathrm{AA}}= -2{\pi}\frac{\sqrt{\mathtt{E}\left(1-\mathtt{E}\right)}}{\left| {\sin \chi } \right|}.
\end{equation}
It is obvious to see that the AA-geometric phase is explicitly affected by the entanglement amount, communicated between the two interacting spins, with a negative sign, meaning that the AA-geometric phase decreases as entanglement increases. This behavior is shown in Figure \ref{amk}(c), it is evident that as the system becomes more entangled, it cumulates a larger negative AA-geometric phase.

This demonstrates a behavior that is quite similar to that observed for the geometric phase \eqref{a19} in the first part, specifically in the range $[0,\mathtt{E}_{\text{c}}]$, and consequently, comparable results can be expected for the AA-geometric phase. Regarding the topological phase generated by the cyclic dynamics of the two-spin system, it is defined as $\gamma_{\operatorname{top}}^{\mathrm{AA}} =-2\pi$. Therefore, it remains independent of entanglement since it represents the AA-geometric phase part that does not take any contribution from the dynamic phase.\par
 
Thereby, we can conclude that the geometric phase fundamentally arises from the curvature and topology of the underlying quantum state space through which the system evolves. Intuitively, as the quantum state vector undergoes parallel transport along a path in this curved manifold, it accumulates a phase that depends not only on the path length but also on how the manifold itself bends and twists. This curvature effectively encodes the geometric constraints imposed by entanglement. When the topology of the state space features structures like the dumbbell shape described, it leads to critical points where the geometric phase behavior changes, as seen at the critical entanglement value $\mathtt{E}_c$. Around this point, the curvature induces a transition in the direction of phase accumulation, from decreasing to increasing, reflecting a shift in how the system's trajectory wraps around the curved space. Thus, the interplay between topology and curvature governs the sign and magnitude of the geometric phase, making it a direct geometric fingerprint of the system's entanglement and evolution path. This highlights that the quantum geometry is not just a mathematical abstraction but a physically meaningful landscape that can be harnessed to control phase-based quantum operations experimentally \cite{c11,c12}.

\subsection{Dynamical facet of quantum entanglement} 
\begin{figure}[htbp]
\begin{center}
\includegraphics[scale=0.3]{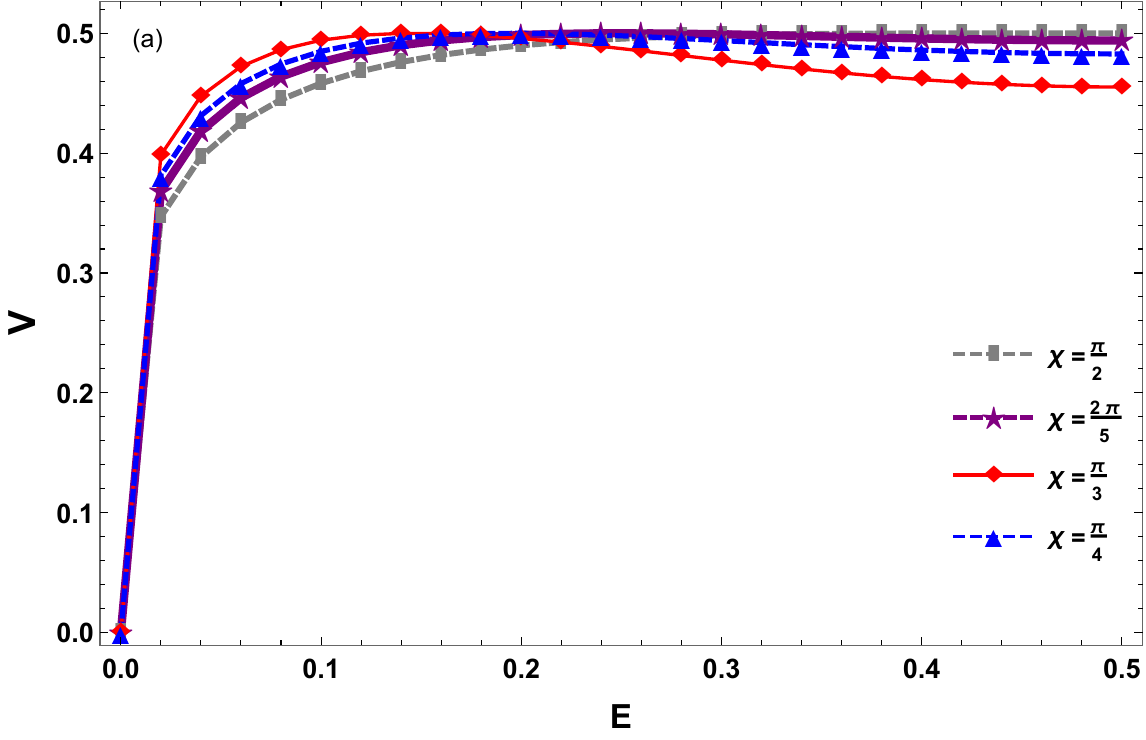}\includegraphics[scale=0.3]{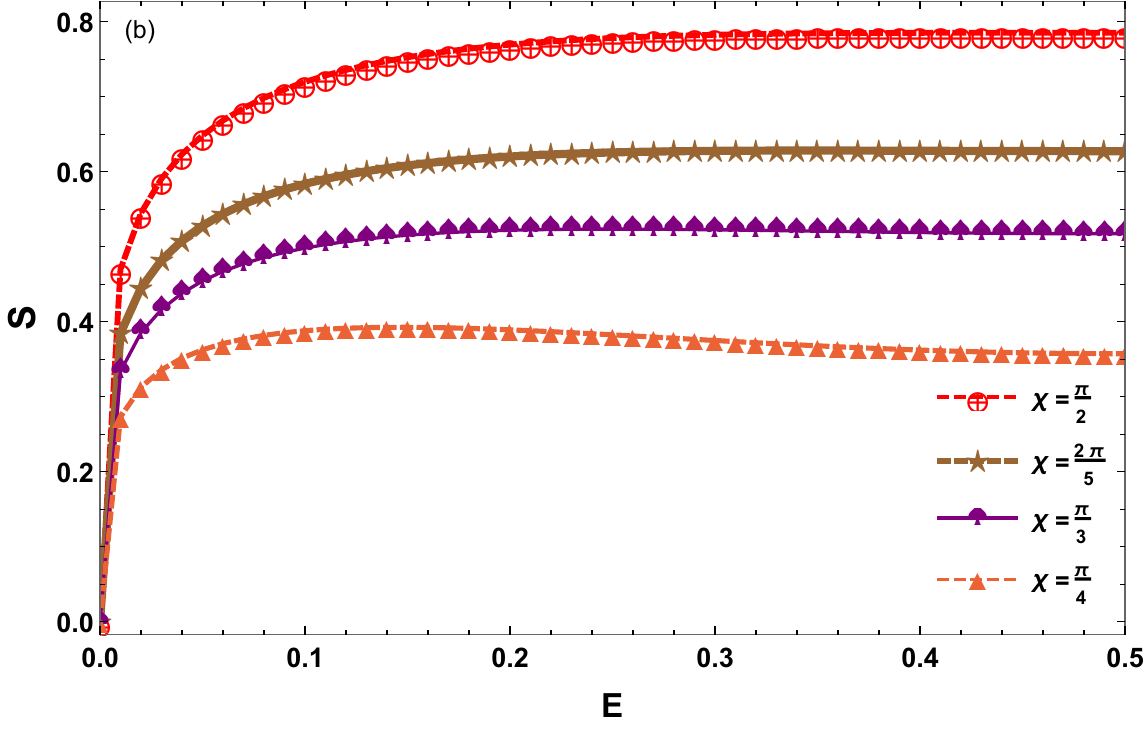}\includegraphics[scale=0.3]{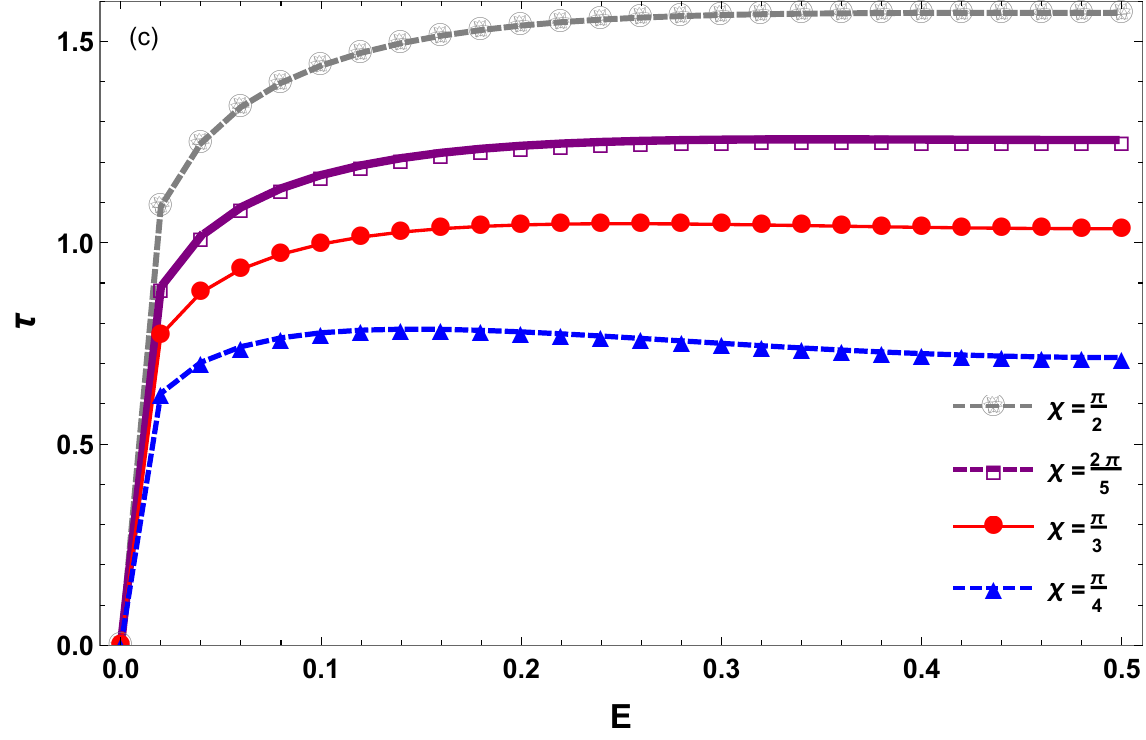}
\caption{The evolution speed [Eq. \eqref{a21}, panel (a)], FS-distance [Eq. \eqref{a22}, panel (b)], and optimal time [Eq. \eqref{a23}, panel (c)]  versus the geometric measure of entanglement \eqref{a13} for some values of $\chi$ by setting $\theta=\pi/2$ and $J=1$.}\label{bhm}
\end{center}
\end{figure}

We will now explore the interconnections between the entanglement between the two spins and the relevant dynamic characteristics, such as the evolution speed and the FS distance that the system traverses along a specific trajectory over the corresponding space of states \eqref{a15}. In addition, we address the problem associated with the quantum brachistochrone utilizing the quantum entanglement contained in the system. Indeed, inserting the result \eqref{a13} within \eqref{a7}, we give the evolution speed in relation with the geometric measure of entanglement under the form
 		\begin{equation}\label{a21}
V = \frac{J}{{\left| {\sin \chi } \right|}}\sqrt {\sqrt {\mathtt{E}(1 - \mathtt{E})} \left( {\left| {\sin \chi } \right| - \sqrt {\mathtt{E}(1 - \mathtt{E})} } \right)}.
 		\end{equation}
Consequently, we manage to connect the speed of the system with its quantum entanglement. Otherwise, the finding \eqref{a21} reveals the obvious link between system evolution and the progression of quantum correlations. Consequently, we conclude that the dynamical properties of such bipartite systems can be revealed by quantifying the quantum correlations involved. For instance, we can govern the rapidity of the system, over the quantum state manifold \eqref{a15}, by adjusting the entanglement level between the two spins. To further explore the influence of entanglement on the system's dynamics, Figure \ref{bhm}(a) displays the behavior of the quantum evolution speed \eqref{a21} with respect to the geometric measure of entanglement. The graph reveals a non-monotonic dependence: initially, as entanglement increases from zero, the speed rises sharply and reaches a maximum value $V_{\text {max }}=J / 2$ at a critical entanglement threshold $\mathrm{E}=\mathrm{E}_{\mathrm{c}}^{\prime}=\sin ^2\left(\frac{\chi}{2}\right)$. In this regime, the presence of quantum correlations effectively enhances the motion, enabling the system to traverse the underlying quantum state manifold \eqref{a15} more rapidly. This behavior is indicative of a locally flattened geometric structure, where the curvature is low and the state space is easier to navigate dynamically.\\
However, beyond this critical point, a further increase in entanglement leads to a gradual decline in speed. This suggests that the geometry of the manifold becomes more curved or constrained, introducing topological features that slow down the evolution. Consequently, while entanglement generally acts as a resource for faster dynamics, excessive correlations may imprint geometric complexities that hinder optimal motion. This dual role of entanglement, as both a facilitator and a limiter of evolution speed, provides a powerful tool for controlling quantum dynamics, with potential applications in implementing time-optimal gates within quantum computing protocols \cite{c13}. From the outcome \eqref{a5}, we demonstrate the FS-distance covered by the system along a specific geodesic length in terms of the geometric measure of entanglement as
 	\begin{equation}\label{a22}
 			S = \frac{\chi}{{\left| {\sin \chi } \right|}}\sqrt {\sqrt {\mathtt{E}(1 - \mathtt{E})} \left( {\left| {\sin \chi } \right| - \sqrt {\mathtt{E}(1 - \mathtt{E})} } \right)}.
 		\end{equation}
 Thus, we reformulate the FS-distance, among any two quantum states within the space of states \eqref{a15}, in relation to the entanglement degree level and the motion time. This accentuates the relevance of empirically analyzing dynamic features, which prompts their use in quantum technology \cite{Sato2012}.\par	

To analyze this result, we depict in Fig. \ref{bhm}(b) the variation of the FS-distance \eqref{a22} against the entanglement degree. We observe that for all values of $\chi$, the FS-distance increases rapidly at low entanglement levels, then saturates as $\mathtt{E}$ approaches $0.5$. Physically, this indicates that moderate entanglement significantly enhances the distinguishability between states, effectively stretching the quantum trajectory through the state space. However, at high entanglement, the FS-distance growth slows down, indicating a saturation effect where additional entanglement no longer contributes proportionally to the system's evolution length. Topologically, this behavior reflects a bounded and smooth manifold structure, where entangled states are embedded within a compact region of state space \eqref{a15}, and the parameter $\chi$ modulates the degree of geometric deformation, higher $\chi$ values result in larger geodesic separations. \par
In simple terms, although entanglement generally helps the system evolve faster and span larger distances in its quantum state space, this benefit doesn’t grow indefinitely. At first, increasing entanglement boosts both speed and reach. But after a certain point, too much entanglement starts to work against the system, slowing things down due to the complex geometry of the quantum space. This subtle balance reminds us that, in designing efficient quantum protocols, it’s not just about maximizing entanglement, it’s about finding the right amount that works best with the geometry of the system. 
At this point, we shall tackle the issue concerning the quantum brachistochrone for the two spin-$1/2$ particles, based on their entanglement level. For this, maximizing the speed \eqref{a21} according to the geometric measure of entanglement, the briefest period necessary to perform the time-optimum motion, over the underlying state space \eqref{a15}, writes
 \begin{equation}\label{a23}
 \boldsymbol{\tau}=\frac{S}{V_{\max}}=\frac{2\chi}{{J\left| {\sin \chi } \right|}}\sqrt {\sqrt {\mathtt{E}(1 - \mathtt{E})} \left( {\left| {\sin \chi } \right| - \sqrt {\mathtt{E}(1 - \mathtt{E})} } \right)}.
\end{equation} 		
This outcome reveals that the optimum time \eqref{a23} is impacted both by the standard time, the coupling constant, and the exchanged entanglement between the two spins. Precisely, we uncover that for $\mathtt{E}=0$  the optimum time nullifies $(\boldsymbol{\tau}=0)$. This is attributed to the fact that the evolved state \eqref{a12p} corresponds to the disentangled beginning state $|\psi_i\rangle=|\frac{1}{2},\frac{1}{2}\rangle$ (no motion case). At the critical entanglement level $\mathtt{E}=\mathtt{E}{\text{c}}^\prime$, the optimal time attains its highest value $(\boldsymbol{\tau}=t)$, denoting that the optimal and standard evolutions of the system coincide. In contrast, for $\mathtt{E}\in \,]0,\mathtt{E}_{\text{c}}^\prime[\, \cup\,  ]\mathtt{E}_{\text{c}}^\prime,1]$  the optimal time is less than the standard time $(\boldsymbol{\tau}<t)$. Further to that, the optimum motion states can be produced through the subsequent transformation
 \begin{equation}
 		\left|\psi_i\right\rangle \rightarrow|\psi(\boldsymbol{\tau})\rangle=\exp({-i \mathrm{H} \boldsymbol{\tau}})\left|\psi_i\right\rangle.
 		\end{equation} 
Moreover, they form one-dimensional manifold of optimum states, within the entire space \eqref{a15}, specified by the FS-metric
 		\begin{equation}\label{a24}
 d {{S}}^2_{\text{opt}}=\frac{1}{{ {\sin^2 \chi }}} {\sqrt {\mathtt{E}(1 - \mathtt{E})} \left( {\left| {\sin \chi } \right| - \sqrt {\mathtt{E}(1 - \mathtt{E})} } \right)}  d{\vartheta}^2,
\end{equation}
so that $\vartheta=J\boldsymbol{\tau}$. The dependence of the optimum time \eqref{a23} on the quantum entanglement is illustrated in Figure \ref{bhm}(c),  we observe that for low entanglement values, the optimal time rises rapidly with increasing $\mathtt{E}$, indicating that initial quantum correlations slow down the system's evolution due to increased curvature in the local geometry of the manifold. As entanglement grows, $\boldsymbol{\tau}$ tends toward saturation, suggesting that the state space becomes increasingly flattened or symmetric, reducing further dynamical resistance. Topologically, this behavior reveals that the manifold is bounded and smooth but deforms with entanglement, adjusting the shape and length of the shortest possible evolution paths (geodesics). Physically, this result reveals a crucial insight: entanglement, while often beneficial, does not linearly accelerate evolution. Instead, the system experiences a non-monotonic speed-time trade-off, where strong entanglement introduces geometric constraints that cap performance gains. Accordingly, we conclude that efficient quantum control requires a balance between entanglement and the geometric properties of the quantum state space \eqref{a15}.
	
 \section{Summary}\label{sec6}		
 In summary, we conducted an in-depth investigation into a physical system consisting of $n$ spin-$1/2$ particles with interactions described by the all-range Ising model. By deriving the metric tensor of the associated quantum state manifold and calculating the corresponding Riemann curvature, we found that the system evolves over a smooth and compact two-dimensional manifold with spherical topology and a dumbbell-like structure. We also investigated the geometric phase accumulated by the system for both arbitrary and cyclic evolutions. Our findings reveal that in the case of arbitrary motion, the geometric phase manifests a nonlinear dependence on time; this distinguishes it from the dynamic phase, which evolves linearly. This discrepancy highlights the intrinsic complexity of the geometric phase in relation to the system's trajectory and the geometry of its state space. Thus, we conclude that the geometric phase can serve as a tool for parameterizing the various potential trajectories of the system. This result holds substantial practical implications for quantum computing. In the cyclic motion case, we calculated the AA-geometric phase and found that it is sensitive to the system's dynamics. However, it depends entirely on the choice of the initial state, revealing its dependence on the geometry of the state manifold rather than on the system's evolution path. Consequently, the cyclic motion trajectories are not parameterizable by the AA-geometric phase. The topological phase arising from this type of evolution is also extensively discussed. By examining the evolution speed and the corresponding FS-distance between quantum states, we resolved the problem related to the quantum brachistochrone. As a result, we revealed that the topology and geometry of the state manifold, shaped by the number of particles $n$, play a crucial role in determining the minimal time for optimal evolutions.\par
On the other hand, by restricting the analysis to two interacting spins, we examined the relevant quantum entanglement from two complementary perspectives. From the geometric perspective, we linked the FS-metric to the geometric measure of entanglement and found that increasing entanglement reduces the curvature of the state space, even driving it to negative values. Highly entangled states thus reside in low-curvature regions, while separable ones lie in high-curvature zones. The geometric phase, examined through the lens of entanglement, captures these geometric constraints: near a critical entanglement value, the curvature induces a reversal in phase accumulation, revealing how topology and geometry govern the phase behavior. Hence, quantum geometry emerges as a tangible structure that encodes entanglement and can be exploited to control phase-based quantum operations. From the dynamical perspective, we related the evolution speed and FS-distance to the geometric measure of entanglement, finding that the existence of quantum correlations enhances quantum evolution only up to a critical threshold. Beyond this point, excessive entanglement induces geometric constraints that slow the dynamics, leading to a non-monotonic speed-time trade-off. The minimal evolution time (brachistochrone time) thus depends nonlinearly on the shared entanglement, highlighting that optimal quantum control arises from a balance between entanglement and the geometry of the state space.

\section*{Acknowledgments:}
Princess Nourah bint Abdulrahman University Researchers Supporting Project number (PNURSP2025R752), Princess Nourah bint Abdulrahman University, Riyadh, Saudi Arabia. This paper also derived from a research grant funded by the Research, Development, and Innovation Authority (RDIA) - Kingdom of Saudi Arabia - with grant number (13325-psu-2023-PSNU-R-3-1-EF). Additionally, the authors would like to thank Prince Sultan University for their support.
\section*{Funding:}
Princess Nourah bint Abdulrahman University Researchers Supporting Project number (PNURSP2025R752), Princess Nourah bint Abdulrahman University, Riyadh, Saudi Arabia.
\section*{Declarations} 

{\bf Data Availability Statement:} The datasets used and/or analysed during the current study available from the corresponding author on reasonable request.\\

{\bf Conflict of interest:} The authors declare that they have no known competing financial interests or personal relationships that could have appeared to influence the work reported in this paper.

\end{document}